\documentclass[Letter]{aa}
\makeatletter
\renewcommand*\aa@pageof{, page \thepage{} of \pageref*{LastPage}}
\makeatother

\usepackage{graphicx}

\usepackage{placeins}

\usepackage{txfonts}

\usepackage{hyperref}

\usepackage{xcolor}
\usepackage[font=small]{caption}

\begin{document}

\title{The odd primordial halo of the Milky Way implied by \textit{Gaia}. A shallow core, but a steep decline}
\titlerunning{Primordial Galactic halo}
\authorrunning{P. Li et al.}

   \author{Pengfei Li
          \inst{1}\thanks{Corresponding author: pli@nju.edu.cn}
          \and
          Stacy S. McGaugh
          \inst{2}
          \and
          Marcel S. Pawlowski\inst{3}
          \and
          Francois Hammer\inst{4}
          \and
          J. A. Sellwood\inst{5}
          }

   \institute{School of Astronomy and Space Science, Nanjing University, Nanjing, Jiangsu 210023, China
              \and
              Department of Astronomy, Case Western Reserve University, 10900 Euclid Avenue, Cleveland, OH 44106, USA
              \and
              Leibniz-Institute for Astrophysics,
              An der Sternwarte 16, 14482 Potsdam, Germany
              \and
              GEPI, Observatoire de Paris, Universit´e PSL, CNRS, Place Jules
Janssen, Meudon, 92195, France.
              \and
            Steward Observatory, University of Arizona, 933 N Cherry Ave, Tucson AZ 85722, USA
                     }

   \date{Received xxx; accepted xxx}
 
  \abstract
{Primordial dark matter halos are well understood from cold dark matter-only simulations. Since they can contract significantly as baryons settle into their centers, direct comparisons with observed galaxies are complicated. We present an approach to reversing the halo contraction by numerically calculating the halo response to baryonic infall and iterating the initial condition. This allowed us to derive spherically averaged primordial dark matter halos for observed galaxies. We applied this approach to the Milky Way and found that the latest \textit{Gaia} measurements for the rotation velocities imply an odd primordial Galactic halo: Its concentration and total mass differ by more than 3$\sigma$ from the predictions, and the density profile presents an inner core that is too shallow and an outer decline that is too steep to be compatible with the cold dark matter paradigm.}

   \maketitle

\section{Introduction}
Disk galaxies have often been used to study the nature of dark matter because their rotation velocities are excellent indicators of their mass distributions \citep[e.g. see][]{Li2020}. With the advent of the European Space Agency cornerstone project \textit{Gaia}, astronomers can measure the parallaxes and proper motions for millions of nearby stars with high accuracy \citep{Vallenari2023}. In this way, it becomes possible to measure the circular Galactic velocity using the 6D phase-space information of individual stars, which cannot be achieved for external galaxies. This makes our Galaxy a unique laboratory for studying dark matter. 

The rotational velocity profile of the Milky Way (MW) has been known to decline at large radii \citep{Eilers2019, Ou2024}, and the latest measurements using \textit{Gaia} DR3 even reported that the decline is Keplerian \citep{Jiao2023, Hammer2024}. This implies that the density of its dark matter halo is extremely low at large radii. The halo density profile is not the same as that of the primordial halo, however, because the latter evolves significantly due to baryonic feedback and compression. It is important to note that the nature of dark matter determines the structure of the pristine halo, which is free of baryonic effects and has been well established with dark matter-only simulations \citep{Navarro1996, Navarro2004}. For massive galaxies such as the Milky Way, studies have shown that stellar feedback has a negligible effect \citep{DiCintio2014, Read2016}, while baryonic compression leads to significant halo contraction \citep{Li2022A&A, Li2022}. The halo contraction can be well modeled and numerically calculated via an adiabatic process \citep{Sellwood2005, Gnedin2011, Piffl2015, Binney2015}. 

We present an approach for inferring the primordial halos by implementing baryonic compression in the Markov chain Monte Carlo (MCMC) approach \citep{Foreman-Mackey2013}. We apply this approach to the Milky Way in this paper.

\section{Numerical modeling and Galactic data}

We used the algorithm proposed by \cite{Young1980} to model the adiabatic contraction of dark matter halos. The main idea was to make use of the conservation of three adiabatic actions (the third action vanishes due to the spherical symmetry, so that only two actions remain, one radial and one azimuthal), and write the distribution function of dark matter particles in terms of the actions, so that they were invariant as the halo contracts. The baryonic potential was then gradually added to the total potential, which led to a gradual change in the density profile. A realization of this algorithm was developed originally as part of the N-body simulation code \textsc{galaxy} \citep{Sellwood2014}, and it was applied to observational data for the first time by \cite{Sellwood2005}. We further developed it to be suitable for diverse observational data and iterable for an MCMC implementation. The MCMC implementation is critical because it allowed us to sample and choose primordial halos based on observed rotation curves. Therefore, we were able to directly link galaxy rotation curves to primordial dark matter halos, which are more comparable to dark matter-only simulations. The code is now made independent as \textsc{compress} and is released at GitHub\footnote{https://github.com/PengfeiLi0606/compress}. The MCMC implementation is also included as an example to show its usage. We provide an in-depth description of the code in Appendix \ref{sec:contraction}.

As an application, we study the primordial halo of the Milky Way in this work. Unlike external galaxies, for which modeling baryonic mass distributions is hampered by the uncertainties on the stellar mass-to-light ratios, our Galaxy has more constraints on the stellar mass, such as the vertical kinematics of local stars \citep{Bovy2013} and microlensing \citep{Wegg2016}. We therefore did not treat baryonic mass distributions as free parameters. Instead, we adopted a wide variety of popular baryonic models from the literature to ensure that our results are robust and independent of the choice of models. In total, we adopted 12 baryonic models (for more details, see Appendix \ref{sec:baryonmodel}).

\begin{figure}
    \centering
    \includegraphics[scale=0.4]{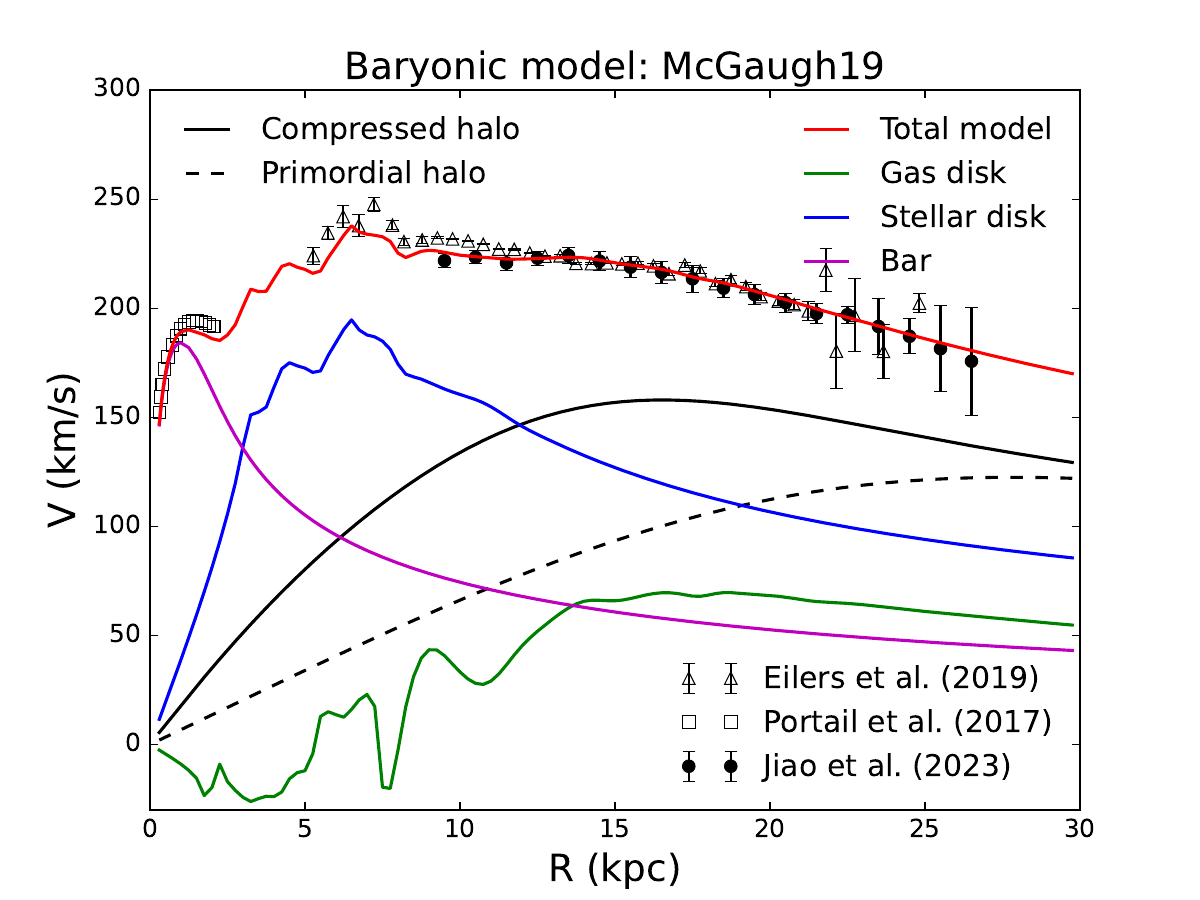}
    \caption{Example of a circular velocity fit using the McGaugh19 model for baryonic mass distributions. The purple, blue, and green lines represent the contributions of the bar, disk, and gas components, respectively. The solid and dashed black lines show the current and primordial dark matter halos, respectively. The solid red line indicates the total velocity profile. The black points show the latest \textit{Gaia} measurements \citep{Jiao2023}, and the gray upward triangles and squares show the terminal velocities from \citep{McClure-Griffiths2007, McClure-Griffiths2016}, and \cite{Portail2017}, respectively. The data marked with open symbols were not fit because they do not consider the systematic uncertainties. The fits with 12 baryonic models are shown in Fig.~\ref{fig:fits}.}
    \label{fig:RCfits}
\end{figure}

The circular Galactic velocity has been extensively measured using the data from \textit{Gaia} DR3 with improved parallaxes and proper motions \citep[e.g.][]{Drimmel2023, Wang2023, Zhou2023, Jiao2023, Ou2024}. The measured velocity profiles from different groups are reasonably consistent with each other, with only small differences at small and large radii. In particular, the measurement by \citet{Jiao2023} considered systematic uncertainties due to the neglected cross-term in the Jeans equation, the uncertainty on the disk scale length, variations in the stellar density profile, and azimuthally varying stellar samples. Their error estimates are therefore relatively more reasonable. We therefore adopted their measurements. The interpretation of these data remains open to debate \citep{Koop2024, Hammer2024, Ou2025}. We investigate here the primordial halo that is implied by the current \textit{Gaia} data when baryonic compression is considered.

To parameterize the primordial halo, we initially used the NFW model \citep{Navarro1996} because it is a prediction of the standard CDM model. However, the NFW model does not fit the Galactic rotation for any baryonic model because the \textit{Gaia} velocity profile declines fast at large radii and is steeper than the outer density slope of the NFW model (see the fits in Fig. \ref{fig:NFWfits}). We therefore employed the more flexible Einasto profile \citep{Einasto1965}, which has an additional shape parameter $\alpha$. \cite{Navarro2004} found that the Einasto profile with a shape parameter $\alpha\sim0.17$ fits their simulated dark matter halos better than the NFW model. This shows that parameterizing primordial halos using the Einasto model can recover simulated halos if the data agree with the CDM model. We used the code \textsc{emcee} \citep{emcee2013} to map the posterior distributions of the primordial halo parameters. The detailed setup is given in Appendix \ref{sec:setup}.

\begin{figure}
    \centering
    \includegraphics[scale=0.4]{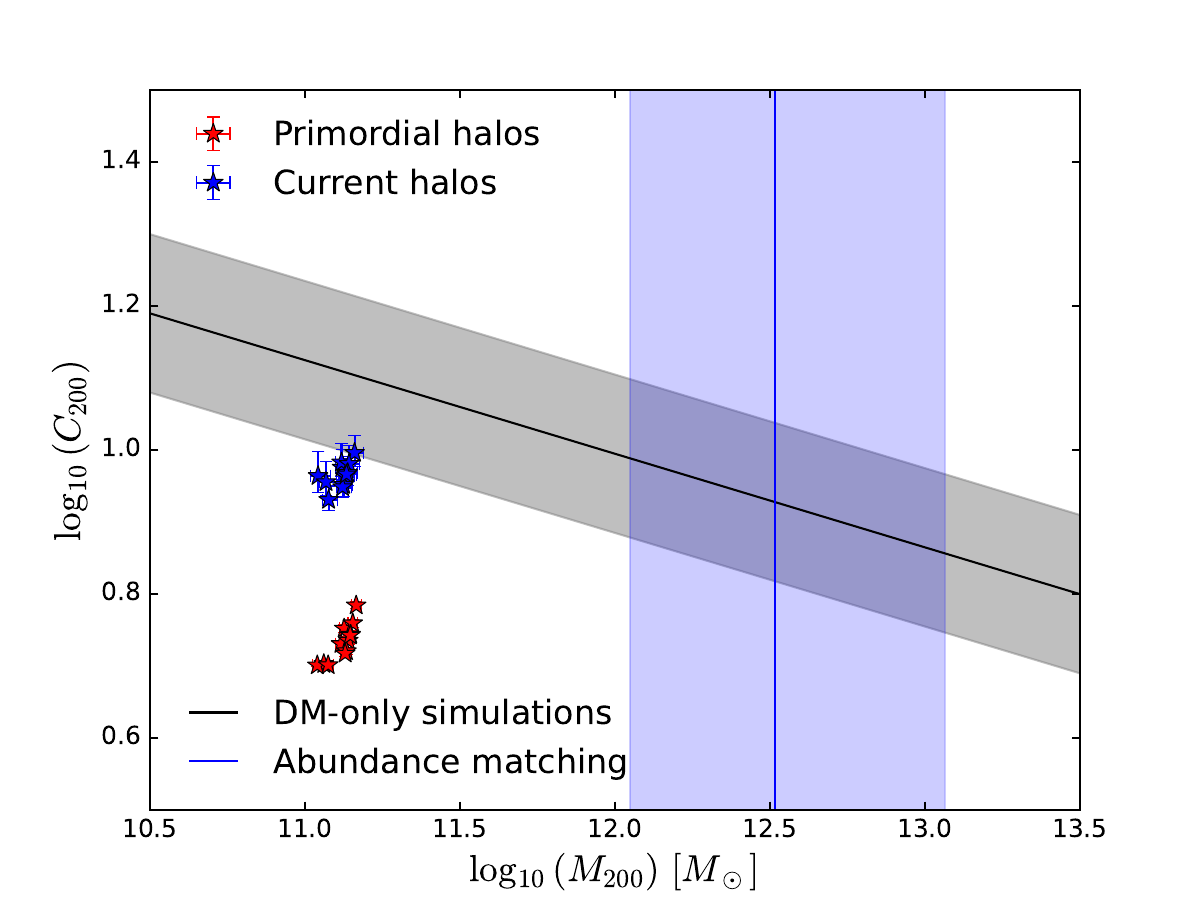}
    \caption{Halo masses and concentrations of the primordial Galactic halos derived from the \textit{Gaia} circular velocity fits using 12 baryonic models. The red and blue stars with errors represent the halos with and without adiabatic contraction, respectively. The predicted halo mass-concentration relation within 1 $\sigma$ from simulations \citep{DuttonMaccio2014} is shown as the declining band. The vertical band shows the expected range of the MW halo mass according to the abundance-matching relation \citep{Moster2013}. The upper and lower limits are set by the highest stellar mass (model A\&S) plus 1 $\sigma$ and the lowest stellar mass (model I) minus 1 $\sigma$, respectively. }
    \label{fig:LCDMpriors}
\end{figure}

\section{Results}

Figure~\ref{fig:RCfits} shows the rotation curve fit using the Einasto model \citep{Einasto1965} for primordial halos and the McGaugh19 model \citep{McGaugh2019} for baryonic mass distributions. For all the adopted baryonic models, the \textit{Gaia} rotation velocity profile fit well with compressed Einasto halos (see Fig.~\ref{fig:RCfits}). The results are summarized in Table \ref{tab:results}. Since the rotation curve data by \cite{Jiao2023} only extend to $\ge$10 kpc, we also plot the inner measurements from \textit{Gaia} DR2 \citep{Eilers2019} and the VVV survey \citep{Portail2017} as a reference, although these data were not fit. The extrapolations of the best Einasto fit can also describe the inner rotation velocity when the McGaugch19 model for baryons is used. This is not always true for other models, however. The primordial and compressed halos contribute rather different rotation velocities. The difference is more profound from 5 to 20 kpc. This shows that the halo structure within 20 kpc changes dramatically after baryonic compression. 

Figure~\ref{fig:LCDMpriors} shows the total halo masses and concentrations of the primordial halos using different baryonic models (also see Table~\ref{tab:results}). For comparison, we also present the halos from the direct fits that did not implement adiabatic contraction. These are simply the current halos and are same as the compressed halos when adiabatic contraction is implemented in our fitting procedure because we fit the same data. Fits without baryonic compression were performed before, and our results are roughly consistent with the results by \cite{Ou2024}. Regardless of whether baryonic compression is included, the results assuming different baryonic models are fairly consistent, which ensures that our results are model independent. Baryonic compression does not change the inferred total halo mass because it only redistributes dark matter particles. The total halo mass ranges from 1.09 to 1.42$\times10^{11}$ M$_\odot$, which is consistent with previous findings \citep{Jiao2023}. They are lower by more than 3$\sigma$ than the predicted values from the abundance-matching relation, however \citep{Moster2013}. This correlation links the predicted halo mass function to observed stellar mass functions. Baryonic compression has a considerable effect on the halo concentration. Neglecting it leads to significantly overestimated halo concentrations. The inferred primordial halos therefore have rather low concentrations, which strongly contradicts the predictions from $N$-body simulations with cold dark matter \citep{DuttonMaccio2014}.

\begin{figure}
    \centering
    \includegraphics[scale=0.4]{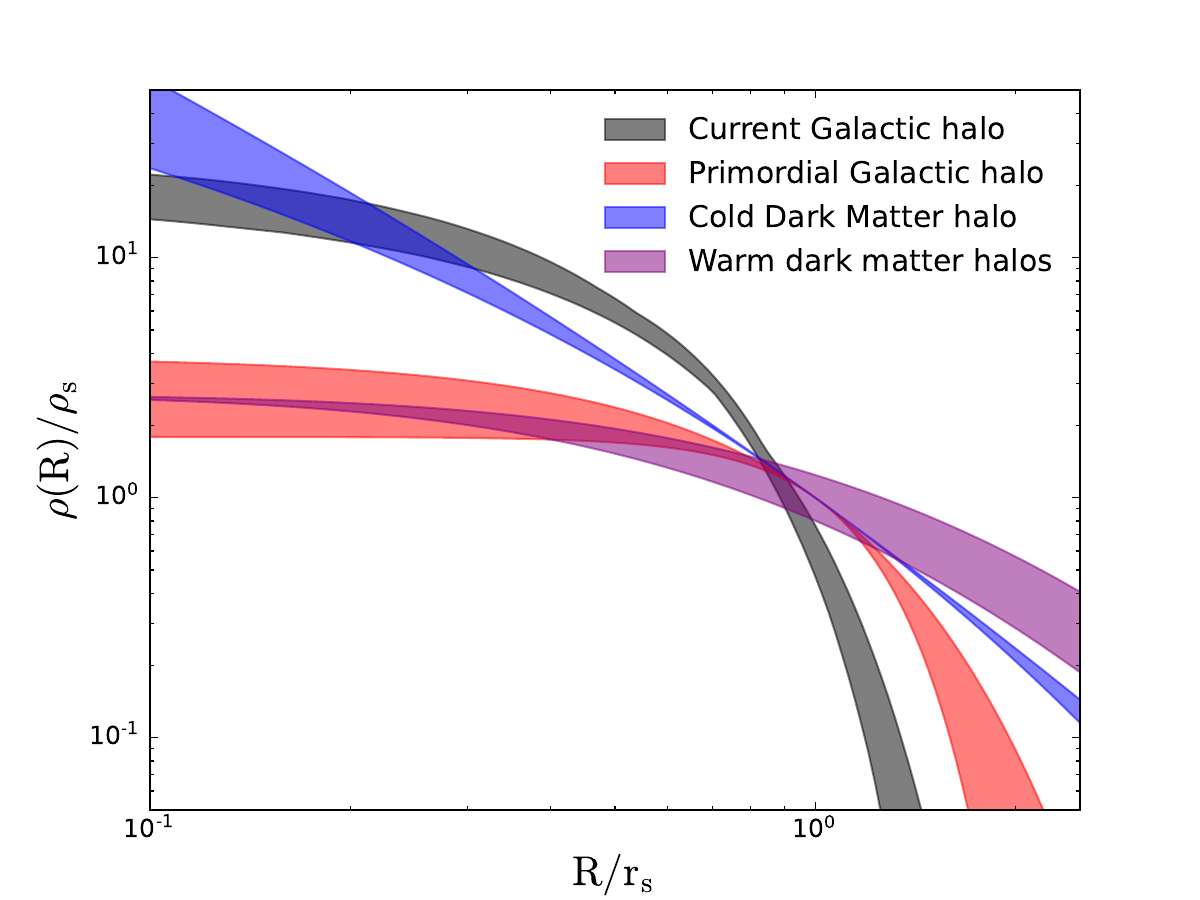}
    \caption{Structure of the inferred primordial and current Galactic halos, along with predictions for the cold and warm dark matter. The density profiles are scaled so that there is no need to assume or consider the masses or concentrations for these halos. The gray band indicates the range of the current halos derived from the \textit{Gaia} velocity fits using the 12 baryonic models, and the red band shows their corresponding primordial halos within 1$\sigma$. The blue band presents the simulated halos with cold dark matter only \citep{DuttonMaccio2014}. The purple band shows the warm dark matter halos (normalized to match the primordial Galactic halo) with a core size spanning from 4.56 kpc (WDM5 in \citealt{Maccio2012}) to 7.0 kpc, corresponding to a particle mass of 0.05 keV and lower.}
    \label{fig:density}
\end{figure}

The structures of primordial and current halos are quite different (Fig.~\ref{fig:density}). Baryonic compression significantly increases the density of the inner halo. Thus, the inner structure of the inferred primordial halos is rather flat. As a trade-off, the dark matter density decreases more quickly at large radii, which slightly helps to explain the quickly declining circular velocity from \textit{Gaia} \citep{Jiao2023}. CDM-only simulations predict a universal cuspy profile, which can be described by the Einasto model with a shape parameter of $\alpha\sim0.17$ \citep{Navarro2004}. In logarithmic space, the density profile of CDM halos is almost linear in the considered radial range (Fig.~\ref{fig:density}). In contrast, the structures of primordial and current halos are apparently curved. Their inner halo structures have a shallow core, which suggests a core-cusp problem. This is a classical problem in dwarf galaxies \citep{Oh2011}. It is usually thought that massive galaxies like the Milky Way host cuspy halos, which is consistent with CDM predictions. The \textit{Gaia} rotation velocities suggest the opposite, however. When baryonic compression is considered, the core of the inferred primordial halo is significantly shallower, which exacerbates the core-cusp problem. The outer density profiles decline steeply as a result from the declining circular velocity profiles. Although including baryonic compression helps to reduce the dark matter density in the outer region, the outer decline of the inferred primordial halo is still too steep (Gaussian-like) compared to that of CDM halos. 

A popular solution for the core-cusp problem is to introduce baryonic feedback, i.e. some physical processes occurring during galaxy formation that drive outflows and help flatten the inner structure of the CDM halo. \citet{Cole2017} showed that a centrally heated halo can have a shallow core in their numerical approach based on distribution functions. The question is whether the amount of energy is sufficient to heat the central dark matter particles. Unlike dwarf galaxies, in which stellar feedback (e.g., supernova explosion and stellar winds) can efficiently transform primordial cusps to cores, our Milky Way is such a massive galaxy that more powerful feedback is required. The only known powerful feedback comes from the accretion of matter onto black holes in galaxy nuclei, known as active galactic nuclei (AGNs; \citealt{Magorrian1998, Silk1998}). AGN feedback has been introduced in the Illustris TNG simulations \citep{Weinberger2017, Pillepich2018a}. It was found, however, that the CDM halos hosted by massive galaxies in TNG have an inner density slope of $\sim$ 2 \citep{Wang2020, Pillepich2024}, which is steeper than that of their primordial counterparts. This is roughly consistent with the finding that only considered baryonic compression \citep{Li2022A&A}. This shows that AGN feedback as implemented in the TNG simulations is not sufficiently powerful to counteract the effect of baryonic compression in the inner region. In addition, the outer halo structure is insensitive to baryonic feedback. Baryon-driven outflows can cause the outer decline to be slightly shallower at most \citep{DiCintio2014}, and this trend is opposite to what is required by the quickly declining rotation velocities. 

Lighter dark matter has also been proposed to generate cored halos, such as warm dark matter (WDM). In~Fig.~\ref{fig:density} we compare our results with the WDM simulations of the MW mass \citep{Maccio2012}. Since the primordial Galactic halo we inferred has a large core, we only show the halo with the lowest particle mass (0.05 keV; i.e., WMD5 in \citealt{Maccio2012}) corresponding to a core size of 4.56 kpc (the lower limit of the purple band), which is the largest core in their simulations, but is still smaller than that of the MW halo. To reproduce the inner structure of the Galactic halo, a core size of $\sim$7 kpc is required (the upper limit of the purple band), but this would also extend the halo more, which contradicts the steep decline suggested by \textit{Gaia}. Moreover, to generate a larger core, the WDM mass would further need to be reduced to below 0.05 keV. As pointed out by \citep{Maccio2012}, a particle mass lower than 0.1 keV would prevent the formation of dwarf galaxies, causing a catch-22 problem.

Fuzzy dark matter \citep{Hu2000} avoids this problem by its nonrelativistic Bose-Einstein condensation \citep{Elgamal2024}, and its quantum stress helps to generate large soliton cores. The inner halo structure might thus be roughly consistent with that of the Milky Way, although we cannot make a robust comparison because our approach only considers gravity, while the halo center is subject to strong quantum stress. In the outer region, however, the halo structure is clearly granular, which is indistinguishable from the CDM halos \citep{Schive2014PRL, Schive2014}. Fuzzy dark matter can therefore hardly reproduce the steep decline of the Galactic halo.

The discrepancies in the inner and outer regions therefore cannot be simultaneously reconciled by simply twisting the nature of dark matter because the structure of the primordial Galactic halo is rather odd: The core is too shallow, and the decline is too steep. Generally speaking, a steep decline suggests that the halo is compact, so that a central cusp is expected, while a shallow core implies that dark matter particles are difficult to aggregate, so that the halo is expected to be extended, corresponding to a slow decline. The shallow core and the steep decline are not expected to be present simultaneously, however, because they imply opposite natures of dark matter. The fact that they do coexist might suggest that some exotic nature of dark matter and more powerful baryonic feedback need to be combined: A steep decline can appear if dark matter particles can gather more closely as a result of some unknown nature; and a shallow core can be generated if the feedback is more powerful. More theoretical and simulation work is required to determine whether it is possible to reproduce the too-shallow-then-too-steep structure.

\section{Discussion and conclusion}
We demonstrated the importance of considering baryonic compression when studying the nature of dark matter with galaxy dynamics. Using an innovative technique, we inferred the primordial dark matter halo of the Milky Way from the \textit{Gaia} rotation curve. \citet{Jiao2023} showed that the declining Galactic rotation curve leads to a significantly lower dark matter halo mass than is expected from the CDM paradigm, more specifically, the abundance-matching relation \citep[e.g.][]{Moster2013, Behroozi2013, Kravtsov2018}. We focused on the detailed structure of the primordial halo, which is directly related to the nature of dark matter. We found that the structure of the primordial Galactic halo is too shallow and then too steep, which is inconsistent with the standard CDM paradigm built from external galaxies and the large-scale structure. If our Milky Way is not special, the nature of dark matter and the model of galaxy formation might need to be adjusted.

The discrepancy at small radii was first indicated by \citet{Binney2015}, who found that an adiabatically compressed NFW halo cannot reconcile various observational constraints, such as microlensing \citep{Popowski2005}. Their focus was on the current configuration of the Milky Way, while we advanced the technique to reverse the adiabatic contraction and thereby infer the primordial halo. \citet{Cautun2020} also investigated the contraction of the Galactic dark matter halo. They selected MW-mass halos from various hydrodynamical and dark matter-only (DMO) simulations. By comparing these halos from the two sets of simulations, they were able to construct a mean mass profile for the compressed halo that was then used to fit the \textit{Gaia} rotation curve \citep{Eilers2019}. Since their simulations assumed cold dark matter, the halos from the DMO simulations are mostly NFW. Initially, we assumed that primordial halos are NFW as well, but we found that the slowly decreasing density profile of NFW halos did not fit the quickly declining \textit{Gaia} rotation curve at large radii. This is also evident from Fig.~10 in \citet{Cautun2020} and similarly in \citet[][see Fig.~ 15]{Binney2023}. We chose a more flexible Einasto profile for primordial halos. For this fundamental reason, they did not find the discrepancy we identified.

It is important to stress that we relied upon the rotation curve deduced from the \textit{Gaia} data. Since different groups found consistent declining Galactic rotation curves using the \textit{Gaia} data, the problem we identified is not specific to the work of \citet{Jiao2023}, but typical to the \textit{Gaia} data. We therefore conclude that these results from \textit{Gaia} differ considerably from the $\Lambda$CDM expectation for our Milky Way. 

\begin{acknowledgements}
We thank Jo Bovy for assistant in using Galpy. P.~L. is supported by the startup funding from Nanjing University. M.~S.~P. acknowledges funding of a Leibniz-Junior Research Group (project number~J94/2020).
\end{acknowledgements}

\bibliographystyle{aa}
\bibliography{PLi}

\begin{appendix}

\section{Method and data}
\subsection{Modeling halo contraction}\label{sec:contraction}

Primordial dark matter halos are self-supported and free of baryonic effects. They start to contract as baryonic gas is accreted into their centers. We modeled the process with the code \textsc{compress}. The code starts with setting up an isolated, extensive and massive spherical initial halo, which must have either a known equilibrium distribution function (DF) or an isotropic one that can be derived by Eddington inversion \citep{Binney2008}. Since the central attraction is changed by embedding a disk and/or bulge at the center of the halo, we need to derive a revised mass profile for the halo in the composite model. Following \citet{Young1980}, we assume that the masses of the disk and bulge were increased adiabatically from zero to their present values, an assumption that does not change the actions of the halo particles \citep{Binney2008}. It was previously assumed angular momentum, one of the actions of an orbit ($J_\phi=2\pi L$), was alone conserved \citep{Blumenthal1986, Klypin2002}, so that the formula would apply only if the orbits of all particles were initially, and remained, precisely circular. The orbits of particles in all reasonable spherical models librate radially, and therefore one must take conservation of radial action ($J_r=2\int^{r_{\rm max}}_{r_{\rm min}}v_r{\rm d}r$ with $v_r$ being the radial velocity) into account when computing the density response to adiabatic changes to the potential. The pressure of radial motions makes a realistic halo more resistant to compression than the naive model with no radial action would predict. Given the conservation of the number of particles and the invariance of adiabatic actions, the DF, when written in terms of actions, is the same after the adiabatic change as before, i.e. $f_0(J_r, J_\phi)=f_n(J_r, J_\phi)$.

As described by \citet{Young1980} and by \citet{Sellwood2005}, a first revised density profile of the halo can be obtained from the original DF by assuming it is an unchanged function of the actions in the new combined potential of the disk and original halo,
\begin{equation}
    \rho'=4\pi\int^{\Phi(\infty)}_{\Phi(r)}\int^{L_{\rm max}}_{0}\frac{Lf(E, L)}{r^2v_r}{\rm d}E{\rm d}L,
\end{equation}
where the distribution function $f(E, L)$ is converted from the invariant $f(J_r, J_\phi=2\pi L)$ using the function $J_r(E, L)$, and the total potential $\Phi(r)$ is updated with the addition of baryonic contribution, i.e. $\Phi=\Phi_0+\Phi_{\rm bar}$. This is the driven force that changes the dark matter density function $\rho(r)$. We then use the first revised density profile of the halo to compute a second revised halo density profile from the DF by again assuming it is an unchanged function of the actions in the new combined potential of the disk and the first revised halo. We iterate in this manner until the change to the halo density between iterations is acceptably small. An initially isotropic DF may become mildly radially biased, which can still be represented by the unchanged actions. Unlike the approximate power-law fit \citep{Gnedin2011}, our procedure gives the detailed mass profile of the compressed halo that results from the adopted initial halo and the observed disk and bulge, which we need to fit the extensive rotation curve.

Note that our procedure assumes the potential remains spherically symmetric, so we must approximate the disk potential by the monopole only term, i.e. using only the disk mass enclosed in a sphere of radius r. From a comparison with a simulation in which a disk was grown slowly inside a spherical halo, the aspherical part of the disk potential caused negligibly small changes to the spherically averaged halo potential. The adiabatic modeling of halo contraction has been tested using N-body simulations in \citep{Sellwood2005}. The spherically averaged density profiles agree remarkably well (see their Fig. 1).

It is important to note that our procedure does take account of adiabatic baryonic feedback, that may rearrange the disk mass through galactic fountains, etc. Of course, explosive feedback that impulsively ejects a significant disk mass fraction is a non-adiabatic change that contradicts our working assumption; such behavior might occur during the assembly of dwarf galaxies, but not those of the mass of the Milky Way.

Note that the adiabatic compression procedure is efficient and runs quite fast, so that we can implement it into the MCMC approach. This way, we randomly sample a large pool of primordial halos with Monte Carlo, and calculate their contracted halos with \textsc{compress}. The contracted halos are then compared to the \textit{Gaia} rotation velocities to build the likelihood function and calculate their probabilities. We then run the Markov chain to generate new pools of primordial halos with larger probabilities, until we obtain a sufficient number of stable pools of primordial halos. Eventually, we select the primordial halo with the highest probability, which will match the \textit{Gaia} circular velocities the best. This implementation improves over the previous approach \citep{Li2022A&A}, which essentially uses a single walker to find the best initial halo. The latter is faster, but it cannot estimate the uncertainties of the fitting parameters. Combining the numerical modeling and the MCMC approach, we can infer the primordial halos from observations.

\subsection{Baryonic models of the Milky Way}\label{sec:baryonmodel}

We adopted 12 baryonic models in this paper, including Model A\&S \citep{Allen1991}, Model I \citep{Pouliasis2017}, Bovy15 \citep{Bovy2015}, McMillan17 \citep{McMillan2017}, B2 \citep{deSalas2019}, and McGaugh19 \citep{McGaugh2019} as detailed below.

{\bf Model A\&S} The classical Model A\&S models a single disk using the Miyamoto–Nagai potential \citep{Miyamoto1975}, given by
\begin{equation}
\label{eq:MN}
    \Phi(R,z) = -\frac{GM}{\sqrt{R^2+\big(a+\sqrt{z^2+b^2}\big)^2}},
\end{equation}
where the parameters are set as $M=8.561\times10^{10}$ M$_\odot$, $a=5.3178$ kpc, $b=0.25$ kpc. The bar is approximated as a Plummer sphere \citep{Plummer1911},
\begin{equation}
\label{eq:Plummer}
    \Phi_{\rm bar}(r) = -\frac{GM_{\rm bar}}{\sqrt{r^2+b^2}},
\end{equation}
where the total mass $M_{\rm bar}=1.406\times10^{10}$ M$_\odot$ and $b=0.3873$ kpc.

{\bf Bovy15} The Bovy15 model also adopts the Miyamoto–Nagai disk (Eq. \ref{eq:MN}) with the total mass $M=6.8\times10^{10}$ M$_\odot$, $a=3.0$ kpc, $b=0.28$ kpc. It models the bar using a power-law density profile with an exponential cutoff,
\begin{equation}
    \rho_{\rm bar}(R, z) = \rho_0\Big(\frac{r_1}{r}\Big)^\alpha\exp{\Big(-\frac{r}{r_c}\Big)^2},
\end{equation}
where the scale density $\rho_0=0.005858$ M$_\odot$ pc$^{-3}$ corresponding to the total bar mass of $0.5\times10^{10}$ M$_\odot$, $\alpha=1.8$, $r_c=1.9$ kpc, and $r_1=8$ kpc. 

{\bf Model I} Model I improves over Model A\&S by including two Miyamoto–Nagai disks (Eq. \ref{eq:MN}), a thin disk with $a_{\rm thin}=5.3$ kpc, $b_{\rm thin}=0.25$ kpc, and a thick disk with $a_{\rm thick}=2.6$ kpc, $b_{\rm thick}=0.8$ kpc. Both have the same total stellar mass $M=3.944\times10^{10}$ M$_\odot$. The Plummer model (Eq. \ref{eq:Plummer}) is also used to approximately describe the bar: $M_{\rm bar}=1.067\times10^{10}$ M$_\odot$ and $b=0.3$ kpc.

{\bf McMillan17} The MCMillan17 model includes thin and thick disks and models their mass density with a double exponential function,
\begin{equation}
\label{eq:doubleexp}
    \rho_{\rm disk}(R,z) = \frac{M}{4\pi HL^2} \exp\Big({-\frac{R}{L}-\frac{|z|}{H}}\Big)
\end{equation}
where $M$, $L$, and $H$ are the total stellar mass, the scale length, and the scale height, respectively. The values that match various observational constraints are: $M_{\rm thin}=3.5\times10^{10}$ M$_\odot$, $L_{\rm thin}=2.50$ kpc, and $H_{\rm thin}=0.3$ kpc for the thin disk; $M_{\rm thick}=1.0\times10^{10}$ M$_\odot$, $L_{\rm thick}=3.02$ kpc, and $H_{\rm thick}=0.9$ kpc for the thick disk. The bar is modeled with an axisymmetric function,
\begin{equation}
    \rho_{\rm bar} = \frac{\rho_0}{(1+r'/r_0)^\alpha}\exp{\Big[-(r'/r_{\rm cut})^2\Big]},\ r'=\sqrt{R^2+(z/q)^2},
\end{equation}
where $\alpha=1.8$, $r_0=0.075$ kpc, $r_{\rm cut}=2.1$ kpc, the axis ratio $q=0.5$, and $\rho_0=98.4$ M$_\odot$ pc$^{-3}$. McMillan17 also models gas distributions using an exponential function with a central hole,
\begin{equation}
    \rho_{\rm gas}(R, z)=\frac{\Sigma_0}{4z_{\rm d}}\exp{\Big(-\frac{R_m}{R}-\frac{R}{R_{\rm d}}\Big)}{\rm sech}^2{(z/2z_{\rm d})}.
\end{equation}
For the H {\footnotesize I} disk, $\Sigma_0=53.1$ M$_\odot$ pc$^{-2}$, $R_{\rm d}=7$ kpc, $R_{\rm m}=4$ kpc, $z_{\rm d}=0.085$ kpc; while for the molecular gas disk, $\Sigma_0=2180$ M$_\odot$ pc$^{-2}$, $R_{\rm d}=1.5$ kpc, $R_{\rm m}=12$ kpc, $z_{\rm d}=0.045$ kpc. The contributions of each component to circular velocity and surface brightness are calculated using the open software GalPot\footnote{https://github.com/PaulMcMillan-Astro/GalPot}.

{\bf B2} The B2 model adopts a single double exponential disk (Eq. \ref{eq:doubleexp}) with $M=3.65\times10^{10}$ M$_\odot$, $L=2.35$ kpc, $H=0.14$ kpc. It also includes components for warm and cold dust, H {\footnotesize I} and molecular gas. These components are modeled with the double exponential model as well, and the parameters are given by
\begin{eqnarray}
    &&M_{\rm cold\ dust} = 7.0\times10^7, L_{\rm cold\ dust}=5.0, H_{\rm cold\ dust}=0.1; \nonumber\\
    &&M_{\rm warm\ dust} = 2.2\times10^5, L_{\rm warm\ dust}=3.3, H_{\rm warm\ dust}=0.09; \nonumber\\
    &&M_{\rm H_2} = 1.3\times10^9, L_{\rm H_2}=2.57, H_{\rm H_2}=0.08; \nonumber\\
    &&M_{\rm HI} = 8.2\times10^9, L_{\rm HI}=18.24, H_{\rm HI}=0.52.
\end{eqnarray} 
The units for mass and scale height/lenth are $M_\odot$ and kpc, respectively. The bar is modeled using the Hernquist potential \citep{Hernquist1990},
\begin{equation}
\label{eq:Hernquist}
    \Phi_{\rm bar} = -\frac{GM_{\rm bar}}{r_b+r},
\end{equation}
where $M_{\rm bar}=1.55\times10^{10}$ M$_\odot$ and $r_b=0.7$ kpc.

{\bf McGaugh19} Unlike previous models, the McGaugh19 model does not assume smooth functions, but numerically maps the azimuthally averaged surface mass densities and the contributions to circular velocity of the stellar disk, the gas disk, and the bar by applying the radial acceleration relation (RAR; \citealt{McGaugh2016PRL, OneLaw}) to measured circular velocities. This approach is based on the fact that the RAR is shown to work in individual disk galaxies \citep{Li2018}. As such, the resulting mass distributions have bumps and wiggles that are typically present in external galaxies.

{\bf Combinations of bar and disk models} Following the literature \citep{Iocco2015, Jiao2021}, we also consider baryonic models from the combinations of bar models and disk models. Two triaxial bar models are considered: E2 \citep{Stanek1997} defined as
\begin{equation}
\label{eq:E2}
    \rho_{\rm bar}(x,y,z) = \rho_0\exp{\Big(-\sqrt{(x/x_b)^2 + (y/y_b)^2 + (z/z_b)^2}\ \Big)},
\end{equation}
where $x_b=0.899$ kpc, $y_b=0.386$ kpc, $z_b=0.250$ kpc, and the scale density is determined from the total mass $M_{\rm bar}=2.41\times10^{10}$ M$_\odot$ using $M_{\rm bar}=8\pi x_by_bz_b\rho_0$; G2 \citep{Stanek1997} given by
\begin{equation}
\label{eq:G2}
    \rho_{\rm bar}(x,y,z) = \rho_0\exp{\Big(-\frac{1}{2}\sqrt{\big(x^2/x_b^2+y^2/y_b^2\big)^2+z^4/z_b^4}\ \Big)},
\end{equation}
where $x_b=1.239$ kpc, $y_b=0.609$ kpc, $z_b=0.438$ kpc, and $\rho_0$ can be calculated from the total mass $M_{\rm bar}=2.12\times10^{10}$ M$_\odot$ via $M_{\rm bar}=6.57252\ \pi x_by_bz_b\rho_0$. Although we adopt triaxial models for the bar, we calculate their azimuthally averaged surface mass density and contributions to circular velocity. This is because the bar is rotating. Averaging azimuthally is equivalent to averaging over time. The disk models are CM \citep{CalchiNovati2011}, J \citep{Juric2008} and dJ \citep{deJong2010}. They all model thin and thick disks using the double exponential function (Eq. \ref{eq:doubleexp}), and the parameters are given as\begin{eqnarray}
    &&{\rm CM\ thin:}\ M = 3.11, L=2.75, H=0.25;\nonumber\\
    &&{\rm CM\ thick:}\ M = 0.82, L=4.10, H=0.75;\\
    &&{\rm J\ thin:}\ M = 3.17, L=2.60, H=0.30;\nonumber\\
    &&{\rm J\ thick:}\ M = 0.90, L=3.60, H=0.90;\\
    &&{\rm dJ\ thin:}\ M = 3.33, L=2.60, H=0.25;\nonumber\\
    &&{\rm dJ\ thick:}\ M = 0.78, L=4.10, H=0.75.
\end{eqnarray}
The units for the total stellar mass ($M$) and the scale length/height ($L$/$H$) are $10^{10}$ M$_\odot$ and kpc, respectively. Combining these bar and disk models, we obtain 6 additional baryonic models: E2CM (E2 bar + CM disk), E2J (E2 bar + J disk), E2dJ (E2 bar + dJ disk), G2CM (G2 bar + CM disk), G2J (G2 bar + J disk), and G2dJ (G2 bar + dJ disk). In total, we adopt 12 baryonic models in this paper. Except for McMillan17 and McGaugh19, we calculate their surface mass density and their contributions to circular velocity using the open Python package galpy\footnote{https://www.galpy.org} \citep{Bovy2015}. Those data are publicly available on GitHub\footnote{https://pengfeili0606.github.io/data}.

\subsection{Halo model and setup}\label{sec:setup}

\begin{table*}
\centering
\caption{Inferred properties of the primordial halos parameterized with the Einasto profile (see Appendix \ref{sec:setup}) before adiabatic contraction assuming a variety of baryonic models (see Appendix \ref{sec:baryonmodel}).}.
\label{tab:results}
\begin{tabular}{lccccc}
\\
\hline
Model & $M_*$ & $M_{200}$ & $C_{200}$ & $r_s$ & $\alpha$\\
& (10$^{10}$$M_\odot$) & (10$^{11}$$M_\odot$) & & (kpc) & \\
\hline
Model A\&S & 9.97 & 1.09$^{+0.05}_{-0.05}$& 5.02$^{+0.03}_{-0.01}$ &19.1& 2.14$^{+0.76}_{-0.66}$ \\
Bovy15 & 7.30 & 1.33$^{+0.05}_{-0.05}$ & 5.63$^{+0.07}_{-0.07}$ &18.2& 2.23$^{+0.69}_{-0.58}$ \\
Model I & 5.01 & 1.20$^{+0.05}_{-0.05}$ & 5.03$^{+0.04}_{-0.02}$ &19.6& 2.40$^{+0.77}_{-0.70}$ \\
McMillan17 & 5.43 & 1.30$^{+0.05}_{-0.05}$ & 5.34$^{+0.07}_{-0.07}$ &18.9& 2.19$^{+0.67}_{-0.57}$ \\
B2 & 5.20 & 1.40$^{+0.05}_{-0.05} $& 5.36$^{+0.06}_{-0.06}$ &17.4& 2.58$^{+0.69}_{-0.66}$ \\
McGaugh19 & 6.16 & 1.19$^{+0.05}_{-0.05}$ & 5.13$^{+0.07}_{-0.06}$ &19.3& 2.75$^{+0.69}_{-0.66}$ \\
E2CM & 6.34 & 1.39$^{+0.05}_{-0.05}$ & 5.47$^{+0.06}_{-0.06}$ &19.1& 2.39$^{+0.69}_{-0.59}$ \\
E2J & 6.48 &1.40$^{+0.05}_{-0.05}$ & 5.36$^{+0.06}_{-0.06}$ &19.6& 2.58$^{+0.69}_{-0.66}$ \\
E2dJ & 6.52 &1.39$^{+0.06}_{-0.05}$ & 5.34$^{+0.06}_{-0.06}$ &19.7& 2.62$^{+0.71}_{-0.64}$ \\
G2CM & 6.05 & 1.41$^{+0.05}_{-0.05}$& 5.69$^{+0.06}_{-0.07}$ &18.2& 2.14$^{+0.68}_{-0.55}$ \\
G2J & 6.19 &1.42$^{+0.05}_{-0.05}$ & 5.56$^{+0.06}_{-0.06}$ &18.8& 2.34$^{+0.65}_{-0.58}$ \\
G2dJ & 6.23 & 1.40$^{+0.06}_{-0.05}$ & 5.54$^{+0.06}_{-0.06}$ &18.8& 2.33$^{+0.69}_{-0.60}$ \\
\hline
\end{tabular}
\end{table*}

We adopt the Einasto model \citep{Einasto1965, Navarro2004} for primordial halos. Its density profile is given by 
\begin{equation}
    \rho_{\rm EIN}(r) = \rho_s\exp\Big\{-\frac{2}{\alpha}\Big[\Big(\frac{r}{r_s}\Big)^\alpha-1\Big]\Big\},
\end{equation}
where the shape parameter $\alpha$, scale density $\rho_s$ and scale radius $r_s$ are free parameters. Its cumulative mass is hence
\begin{equation}
    M_{\rm EIN}(r) = 4\pi\rho_sr_s^3\frac{1}{\alpha}\exp\Big(\frac{2}{\alpha}\Big)\Big(\frac{2}{\alpha}\Big)^{-\frac{3}{\alpha}}\Gamma\Big(\frac{3}{\alpha}, \frac{2}{\alpha}x^{\alpha}\Big),
\end{equation}
where $x=r/r_s$, $\Gamma(a,x)=\int^x_0t^{a-1}e^{-t}{\rm d}t$ is the incomplete Gamma function. The total halo mass $M_{200}$ and concentration $C_{200}$ are defined as
\begin{equation}
    GM_{200} = V^2_{200}r_{200};\ C_{200} = r_{200}/r_s,
\end{equation}
where $r_{200}$ is the radius that encloses a mean halo density of 200 times the critical density of the Universe, and $V_{200}$ is the rotation velocity due to the dark matter halo at $r_{200}$. The velocity profile is then given by
\begin{equation}
    \frac{V_{\rm EIN}(r)}{V_{200}} = \sqrt{\frac{C_{200}}{x}\frac{\Gamma(\frac{3}{\alpha}, \frac{2}{\alpha}x^{\alpha})}{\Gamma(\frac{3}{\alpha}, \frac{2}{\alpha}C_{200}^{\alpha})}}
\end{equation}
The total velocity is the summation of baryonic contributions and the dark matter contribution,
\begin{equation}
    V_{\rm tot}^2 = V_{\rm bar}^2 + V_{\rm disk}^2 + V_{\rm gas}^2 + V_{\rm EIN,compress}^2,
\end{equation}
where $V_{\rm bar}$, $V_{\rm disk}$, and $V_{\rm gas}$ are the rotation velocity contributed by the bar, the disk, and the gas (including dusts) components, respectively; $V_{\rm EIN,compress}$ is the contribution from the compressed Einasto halo. Those baryonic components are fixed for each given baryonic model, and we consider 12 baryonic models from the literature as described in \ref{sec:baryonmodel}. Therefore, the free parameters are only on the halo, i.e. $V_{200}$, $C_{200}$, and $\alpha$.

\begin{figure*}[b]
\centering
    \includegraphics[scale=0.25]{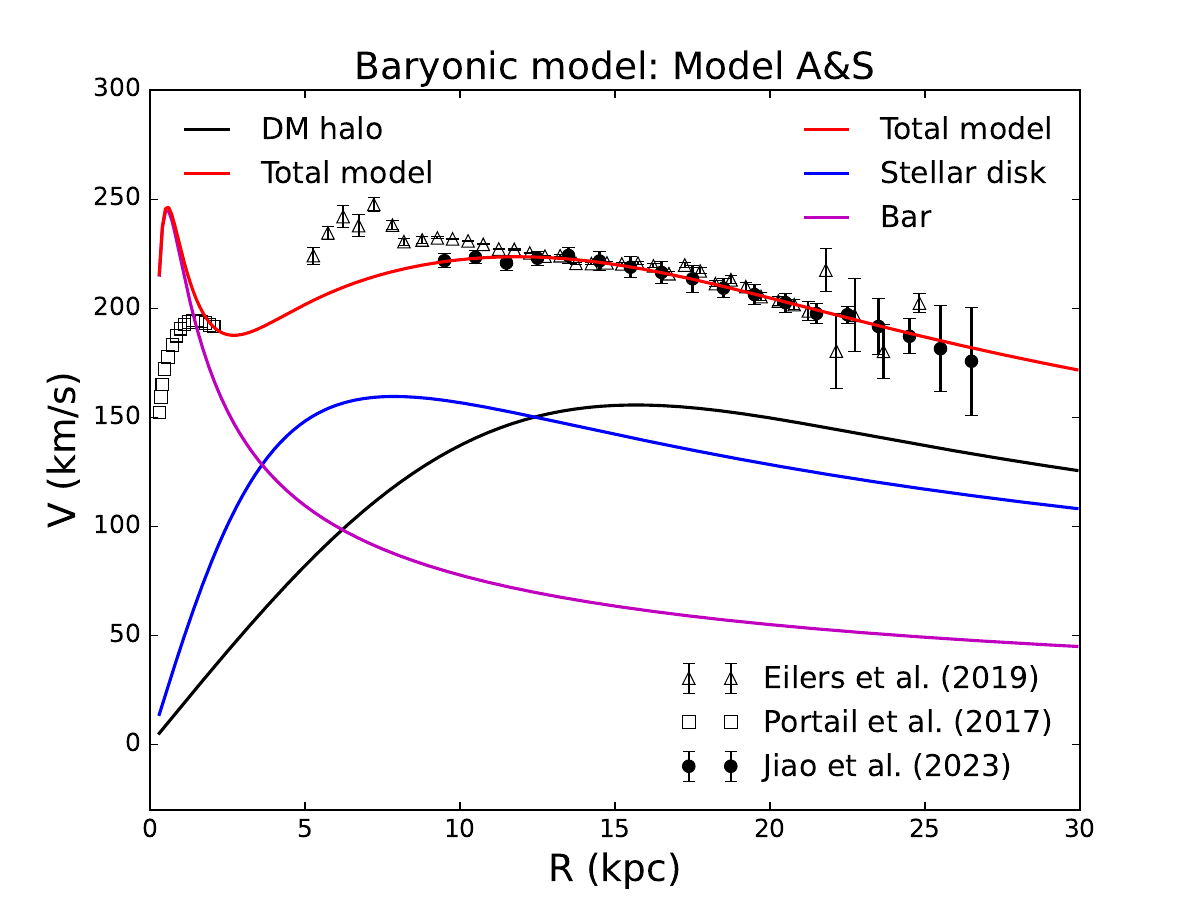}\includegraphics[scale=0.18]{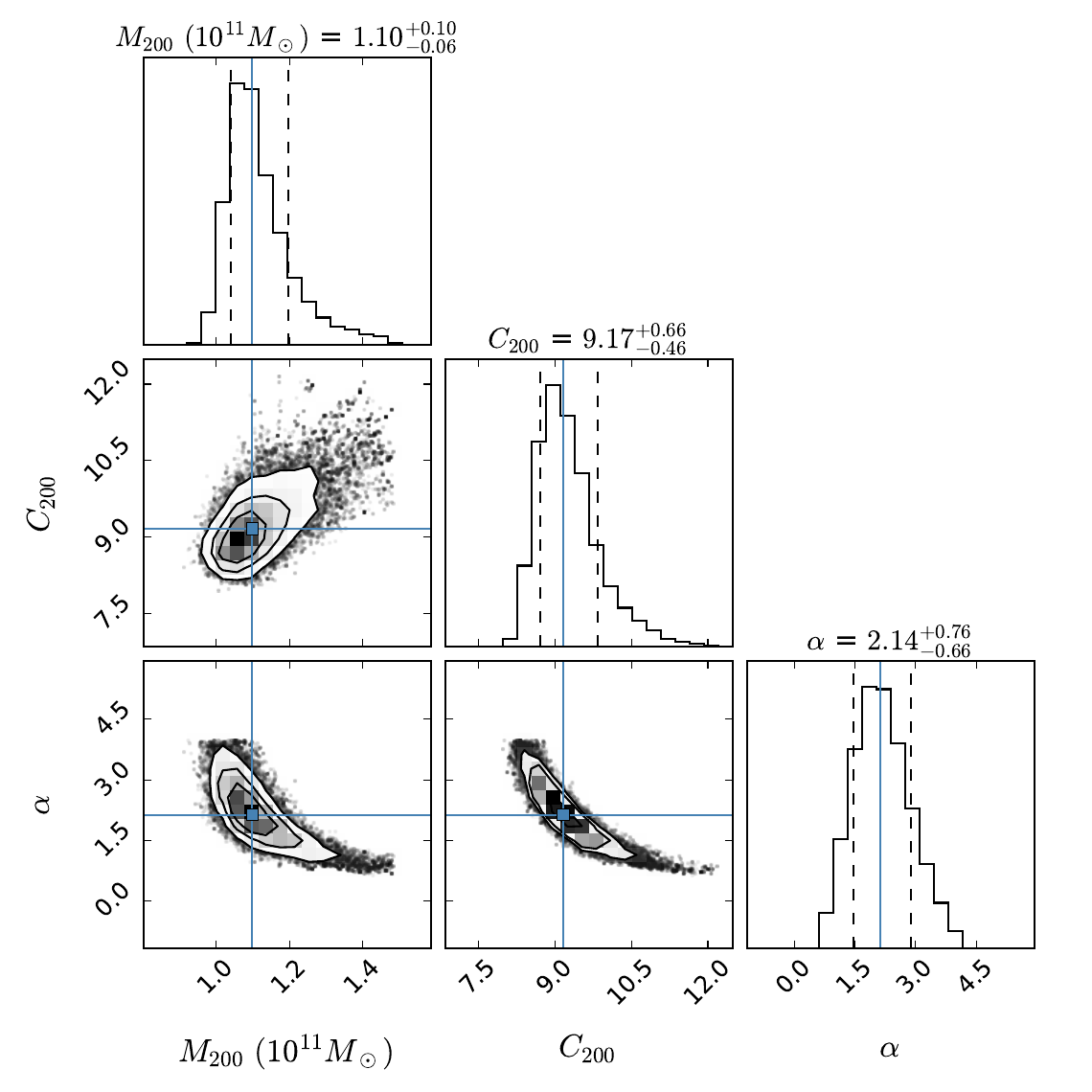}
    \includegraphics[scale=0.25]{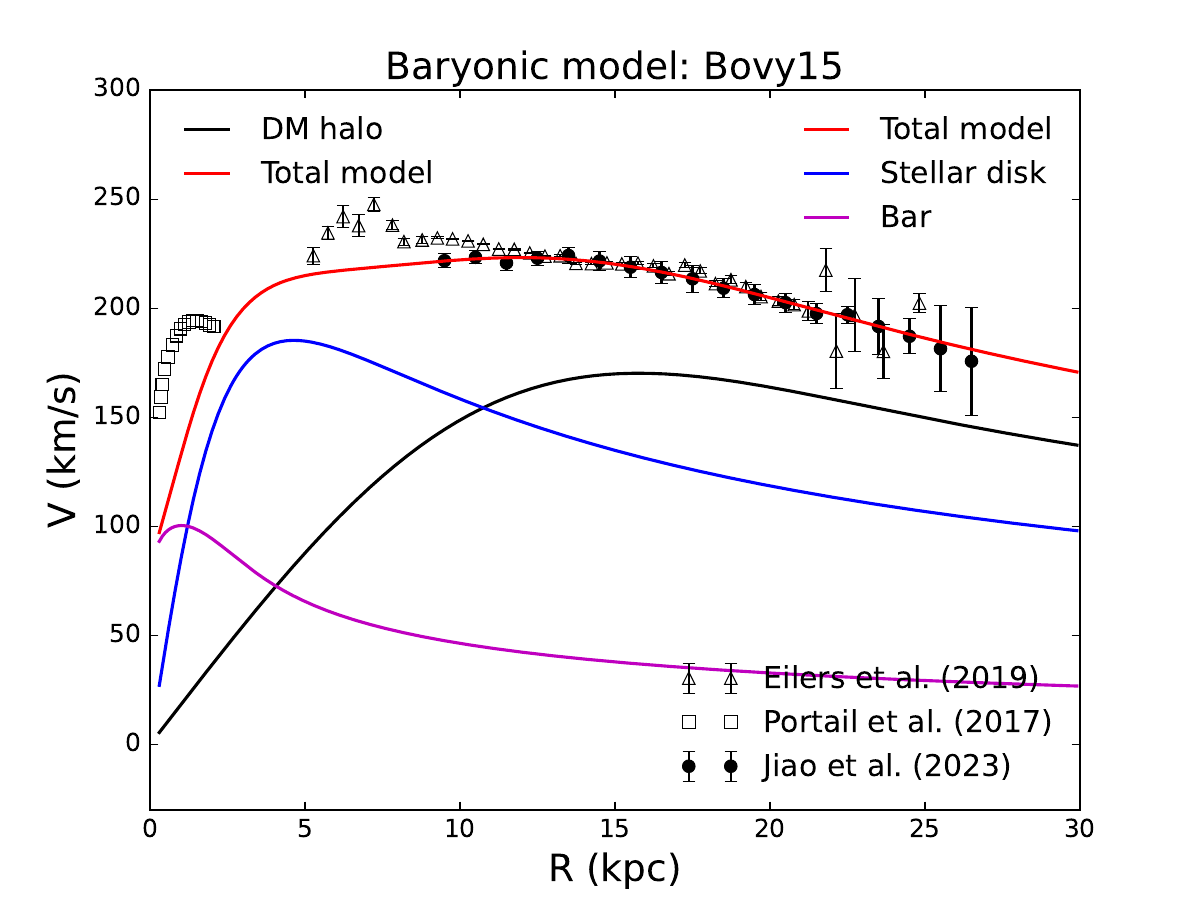}\includegraphics[scale=0.18]{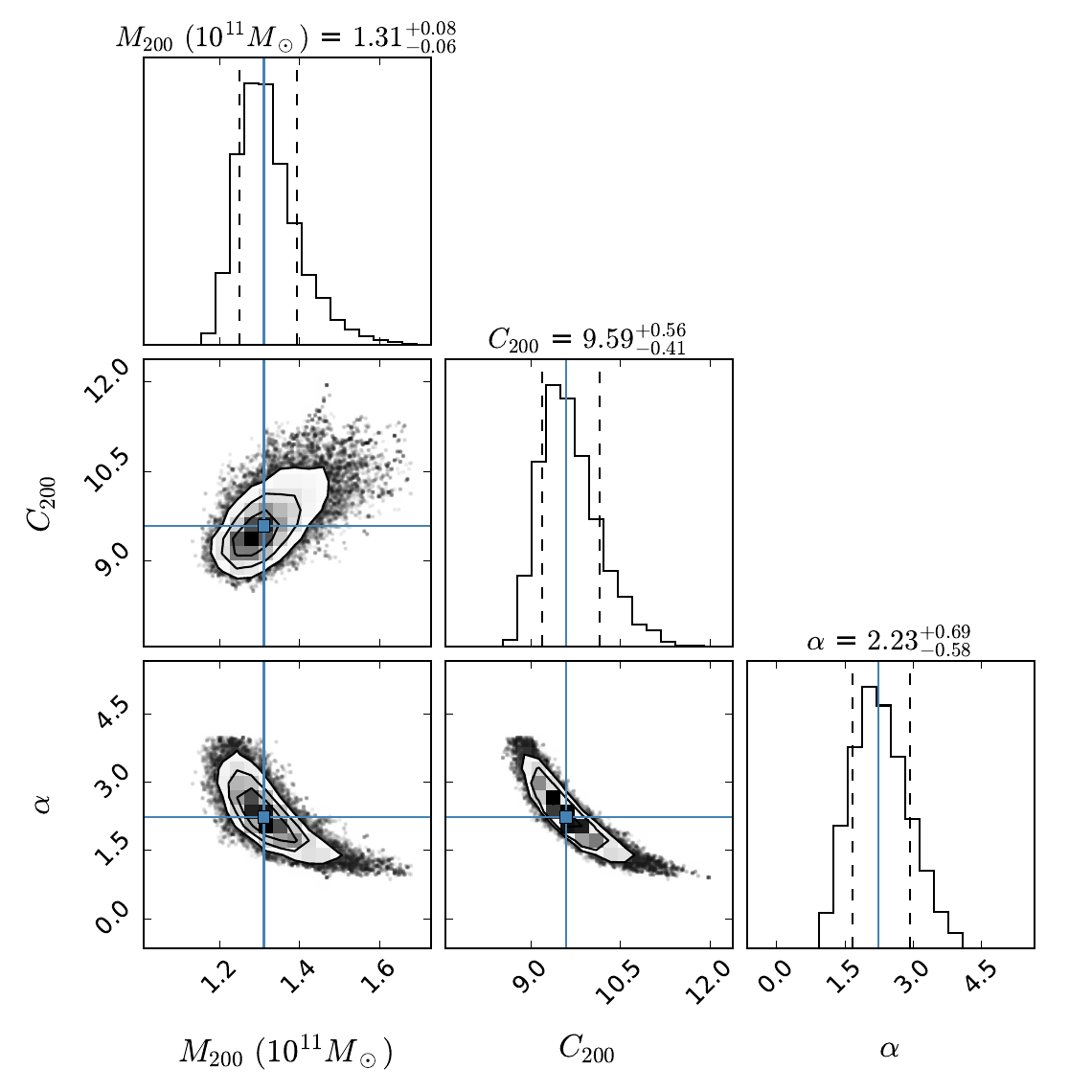}\\
    \includegraphics[scale=0.25]{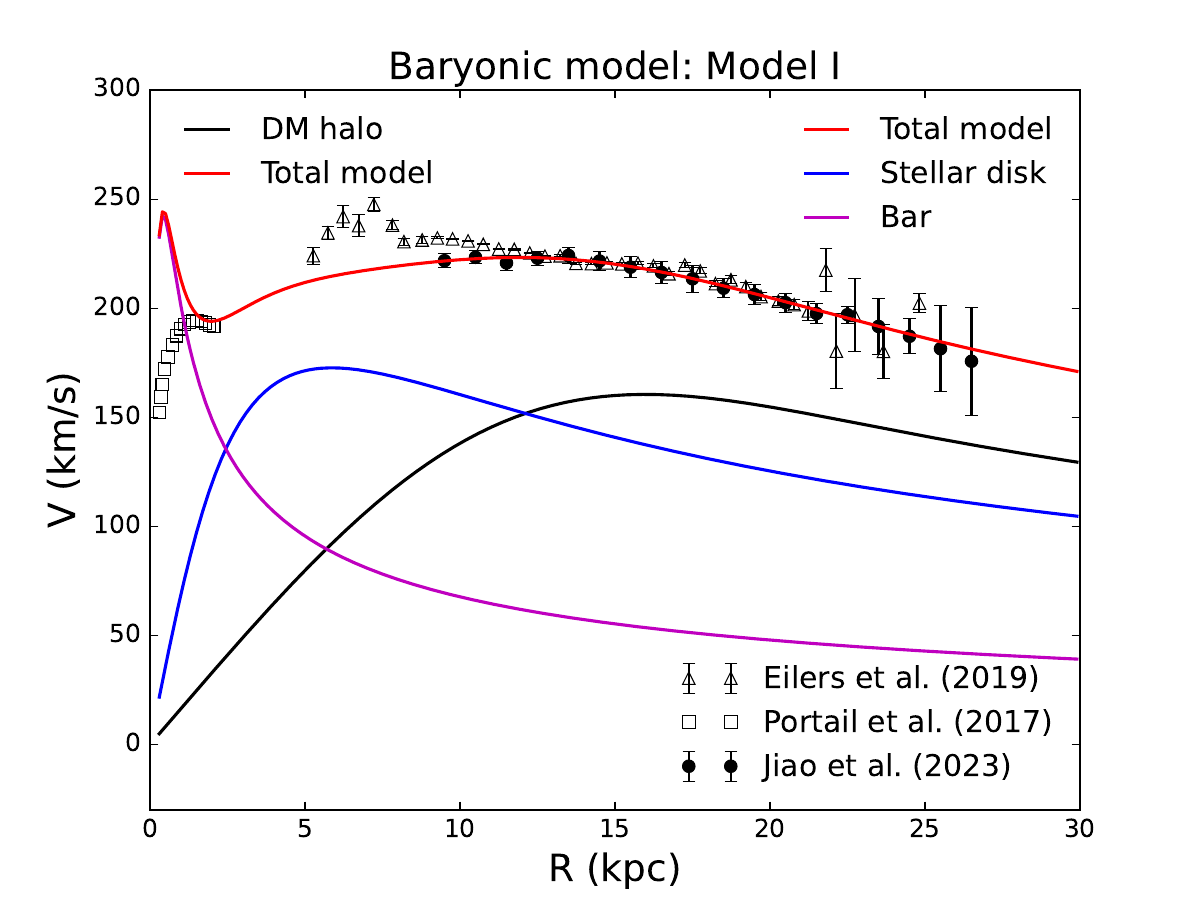}\includegraphics[scale=0.18]{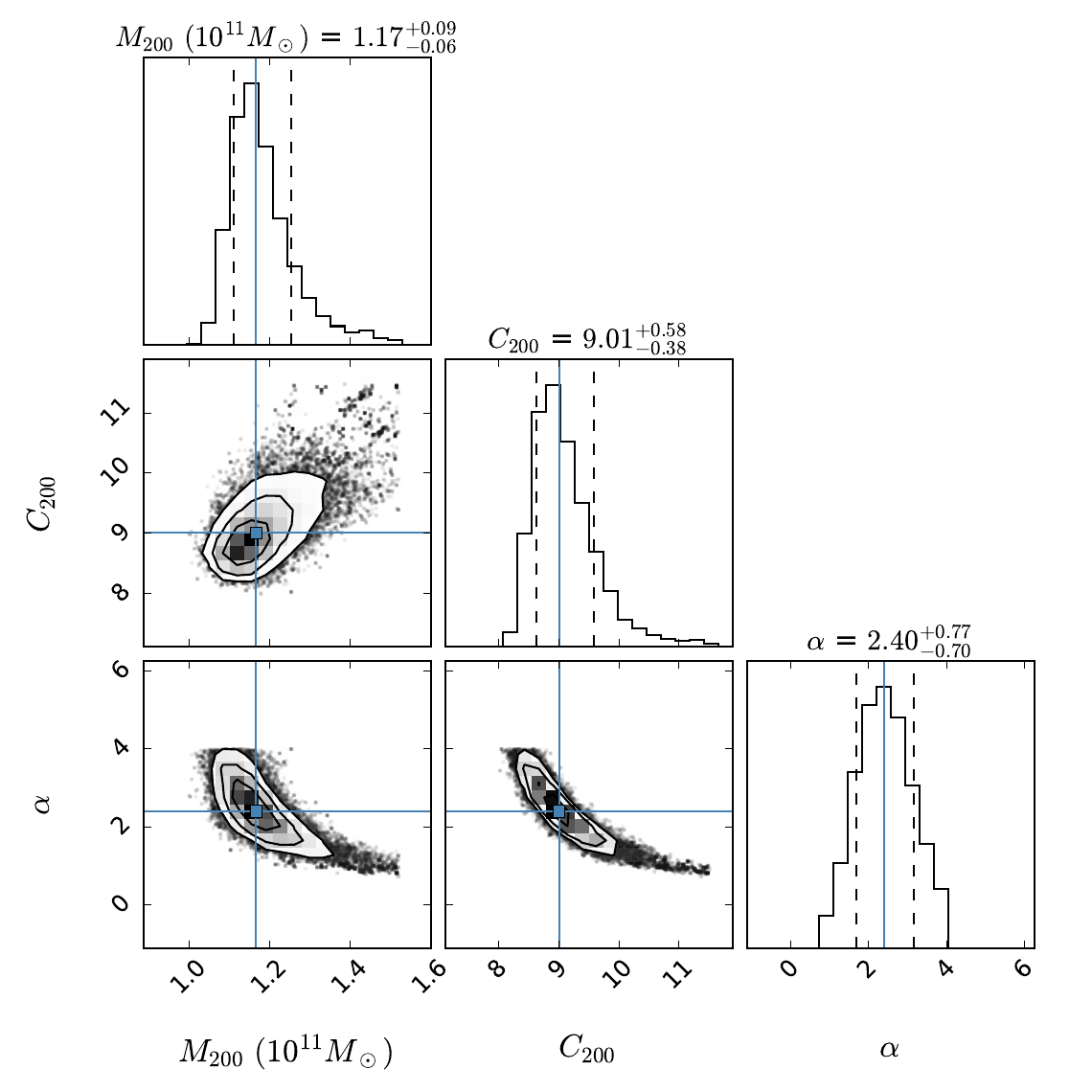}
    \includegraphics[scale=0.25]{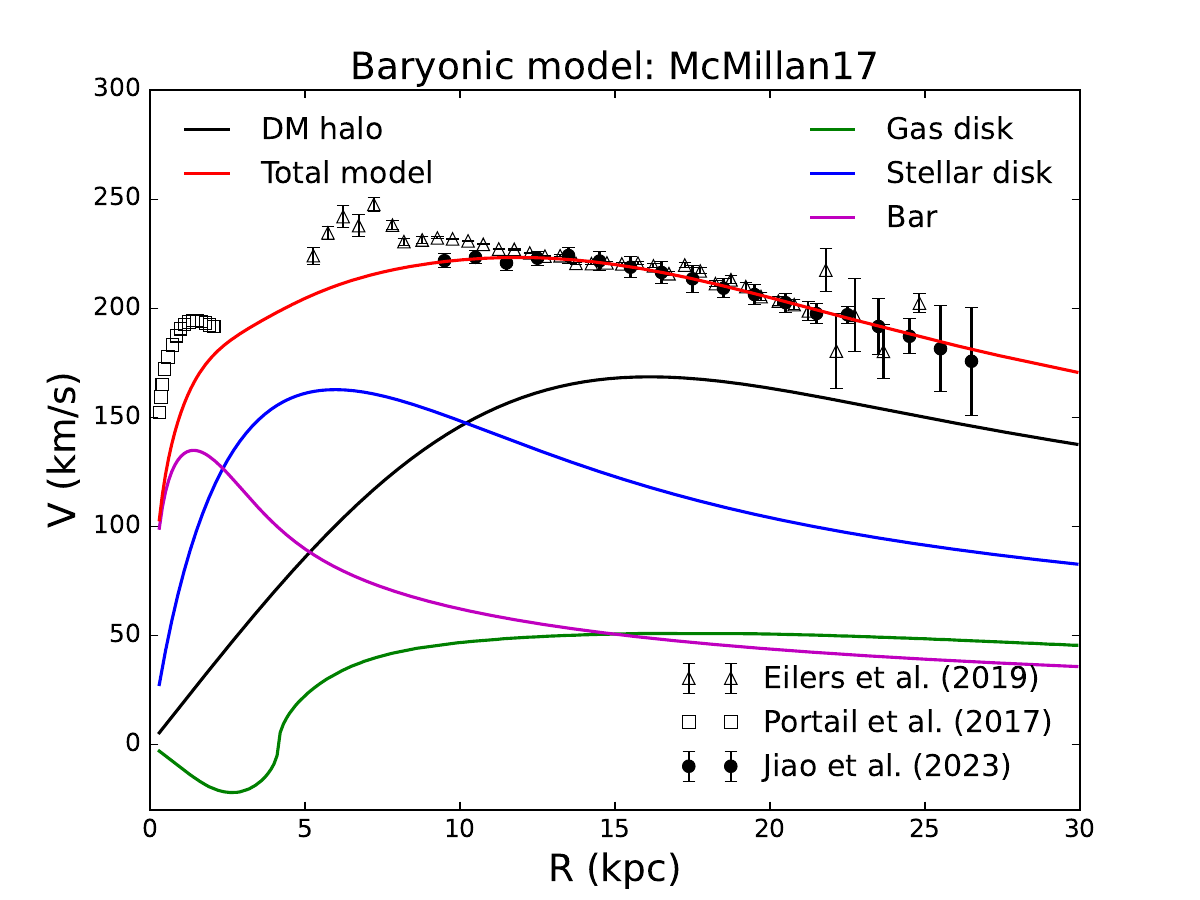}\includegraphics[scale=0.18]{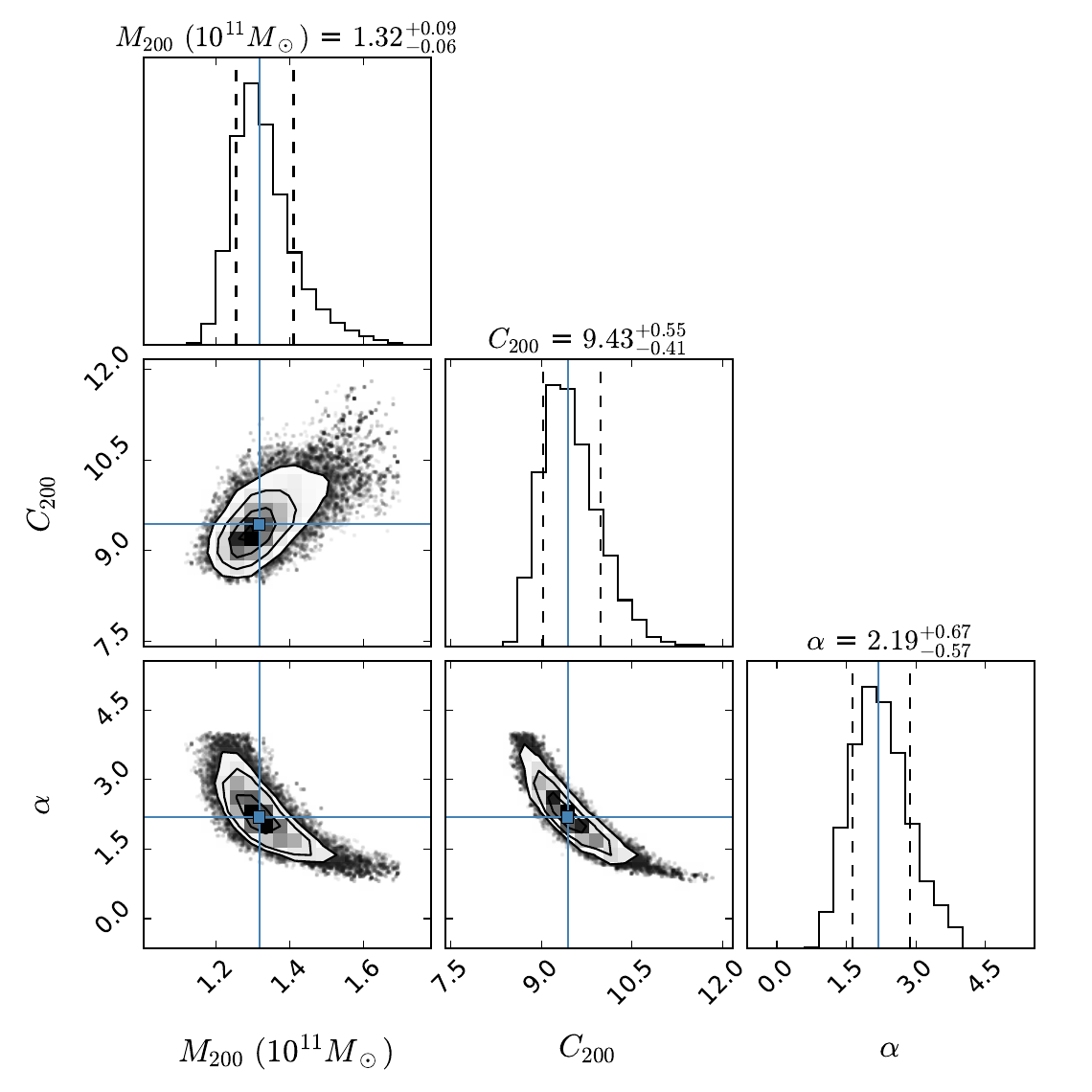}\\
    \includegraphics[scale=0.25]{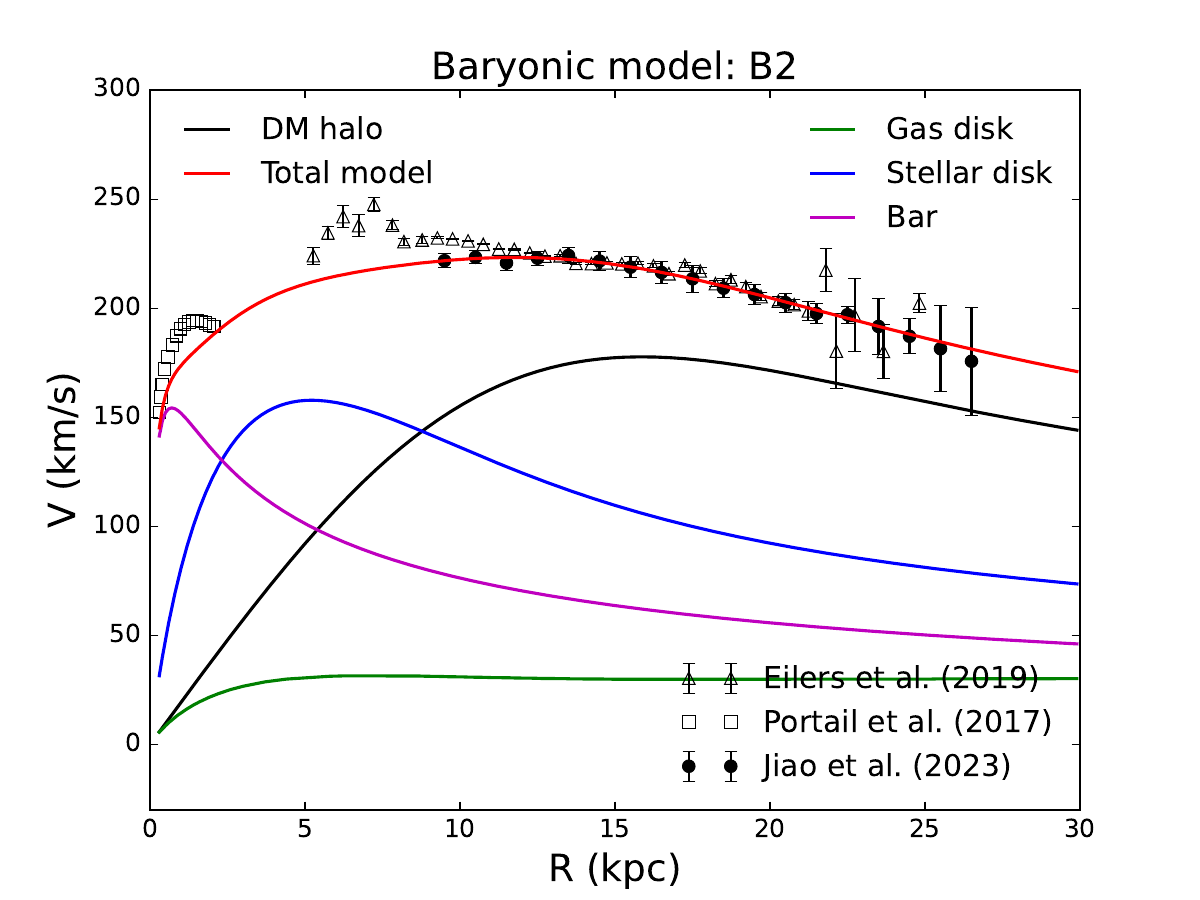}\includegraphics[scale=0.18]{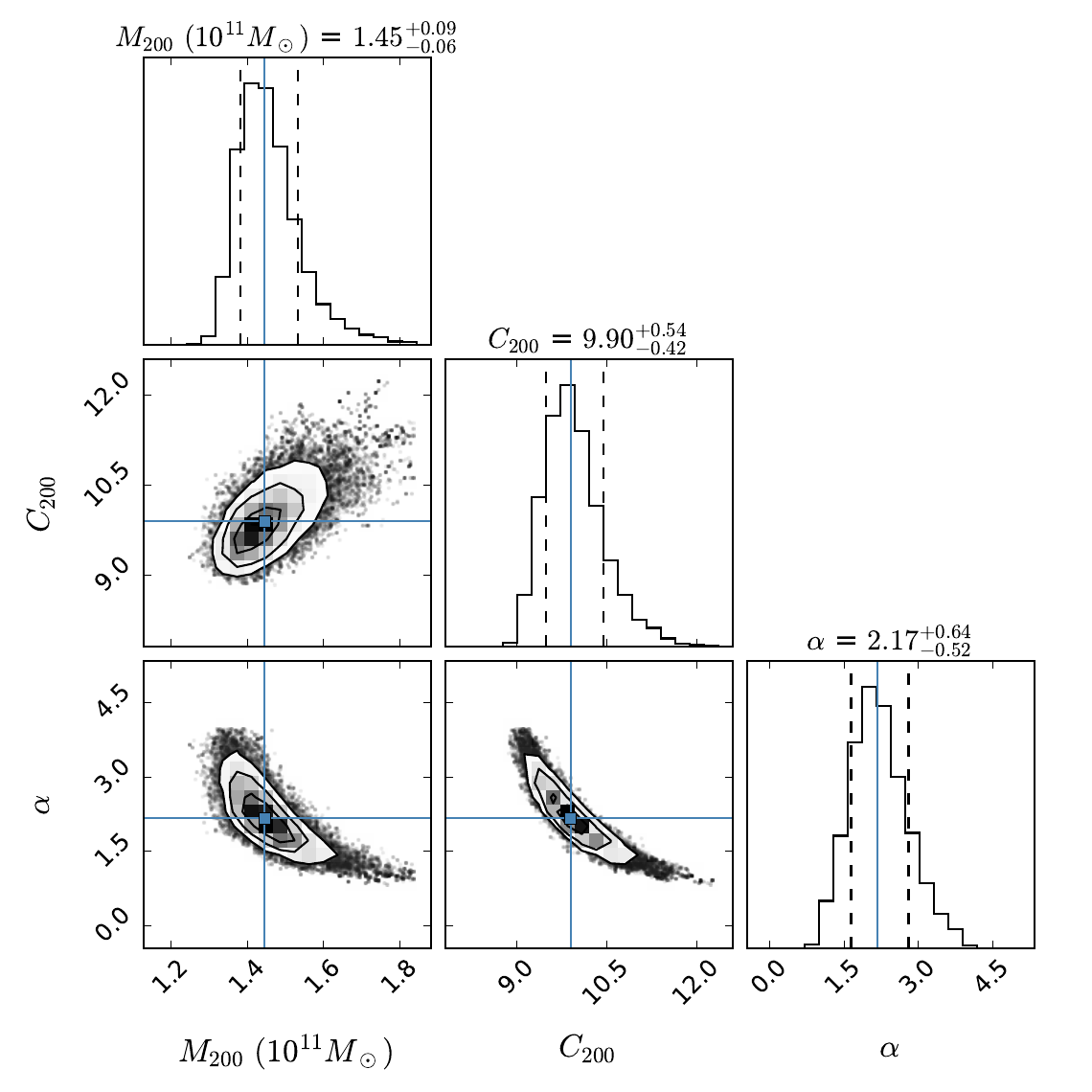}
    \includegraphics[scale=0.25]{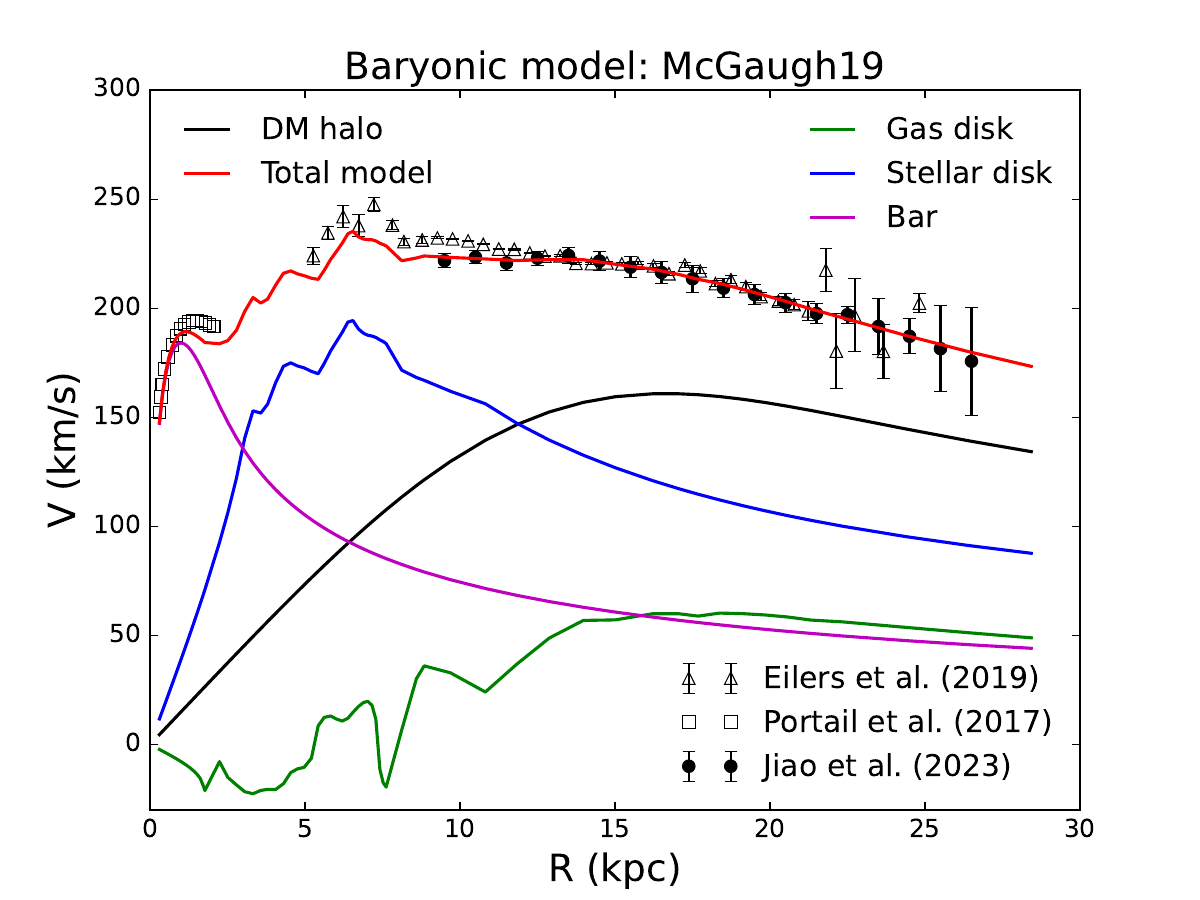}\includegraphics[scale=0.18]{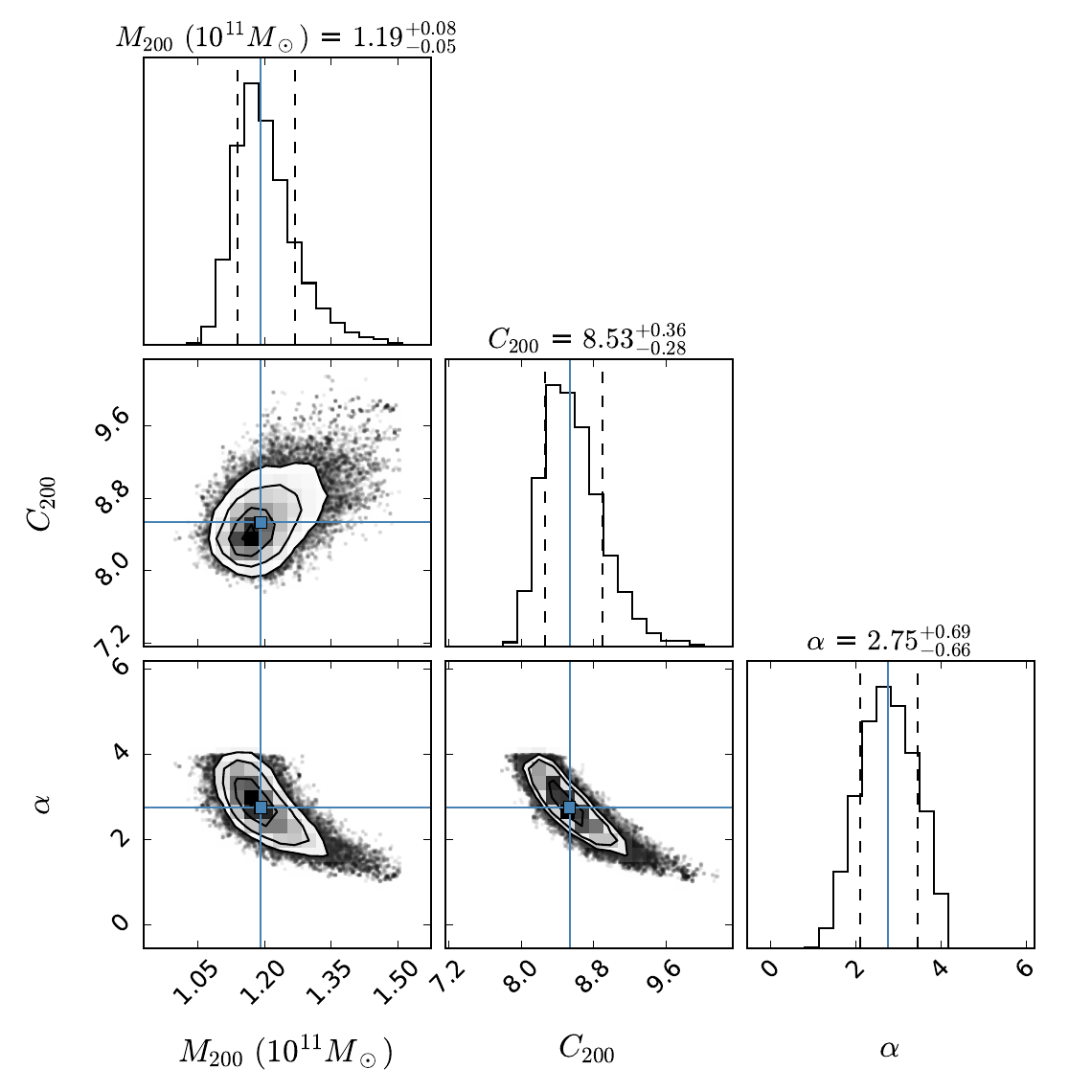}\\
    \includegraphics[scale=0.25]{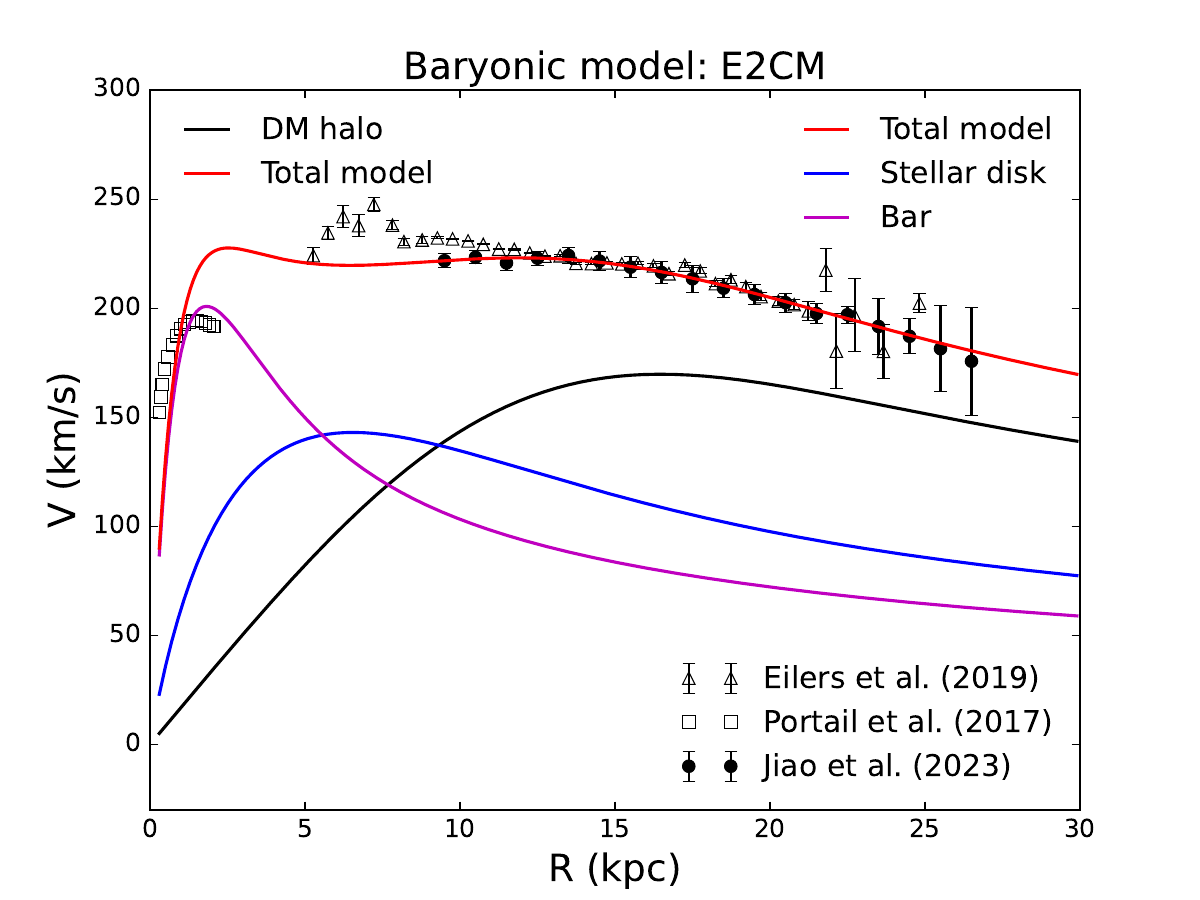}\includegraphics[scale=0.18]{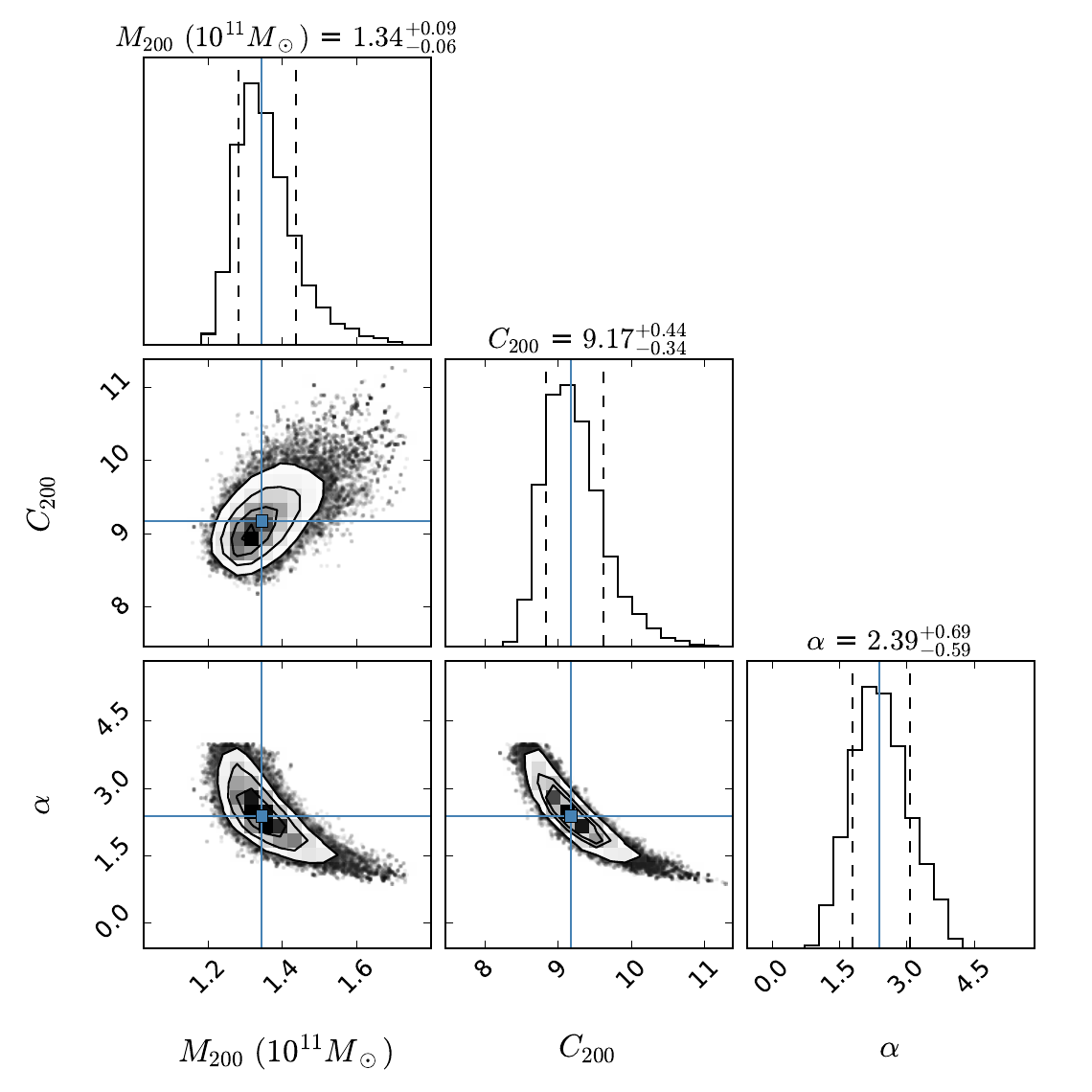}
    \includegraphics[scale=0.25]{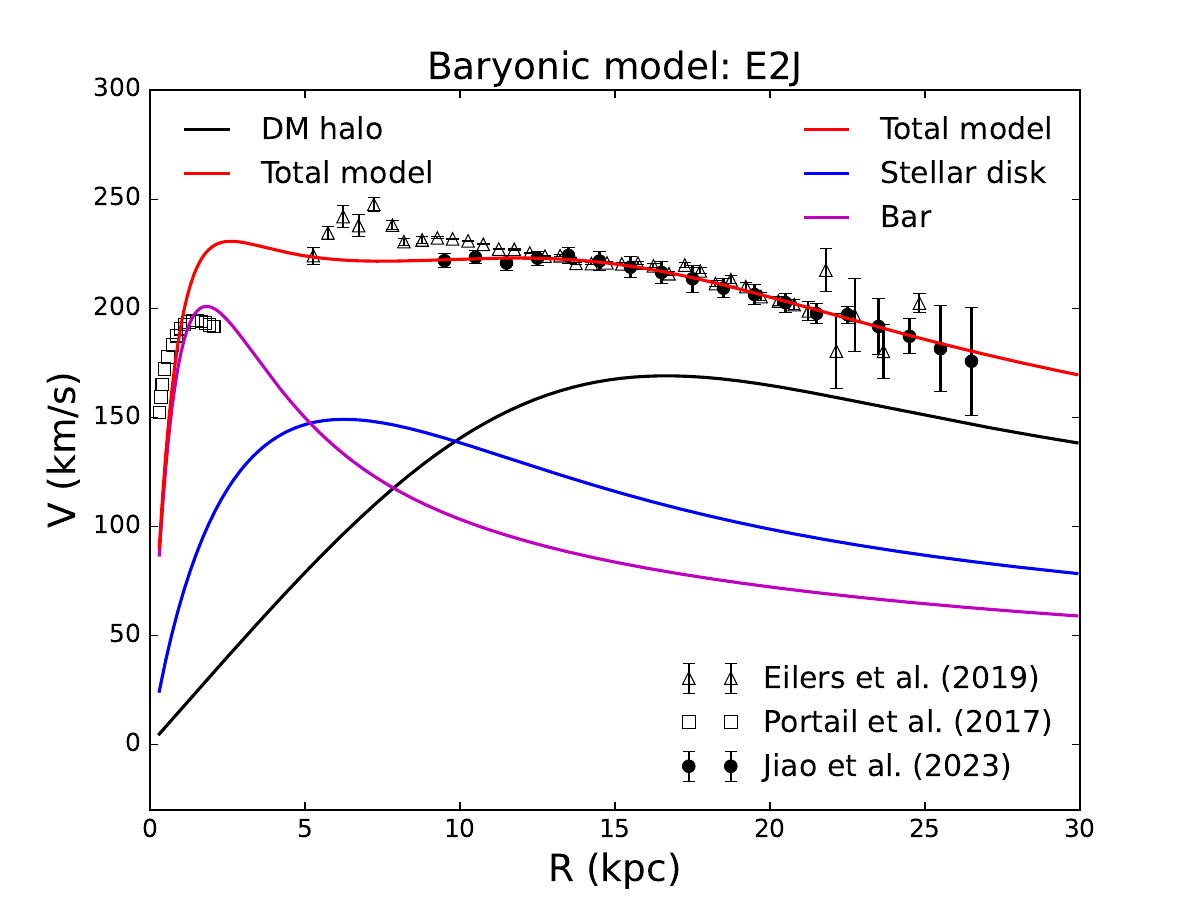}\includegraphics[scale=0.18]{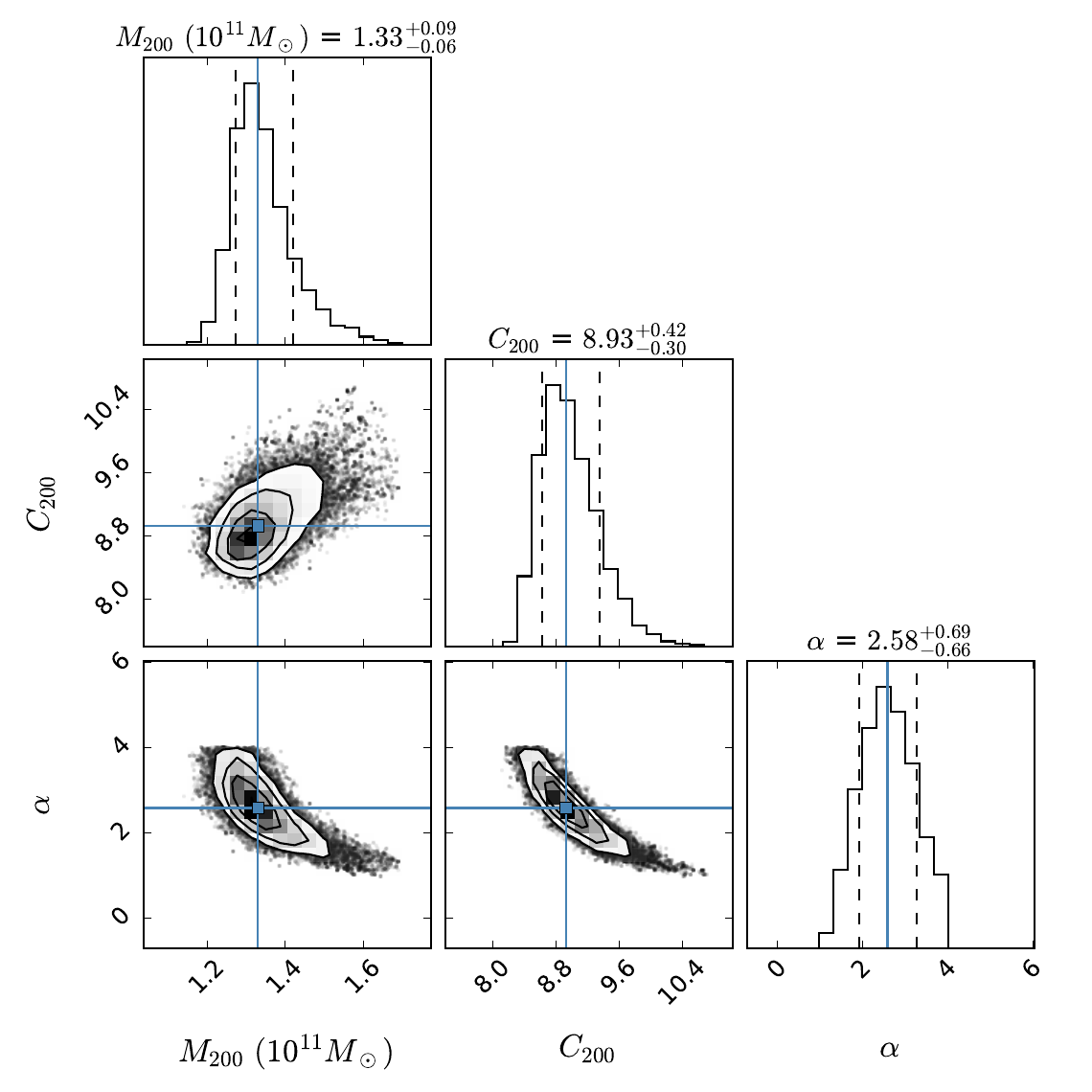}\\
    \includegraphics[scale=0.25]{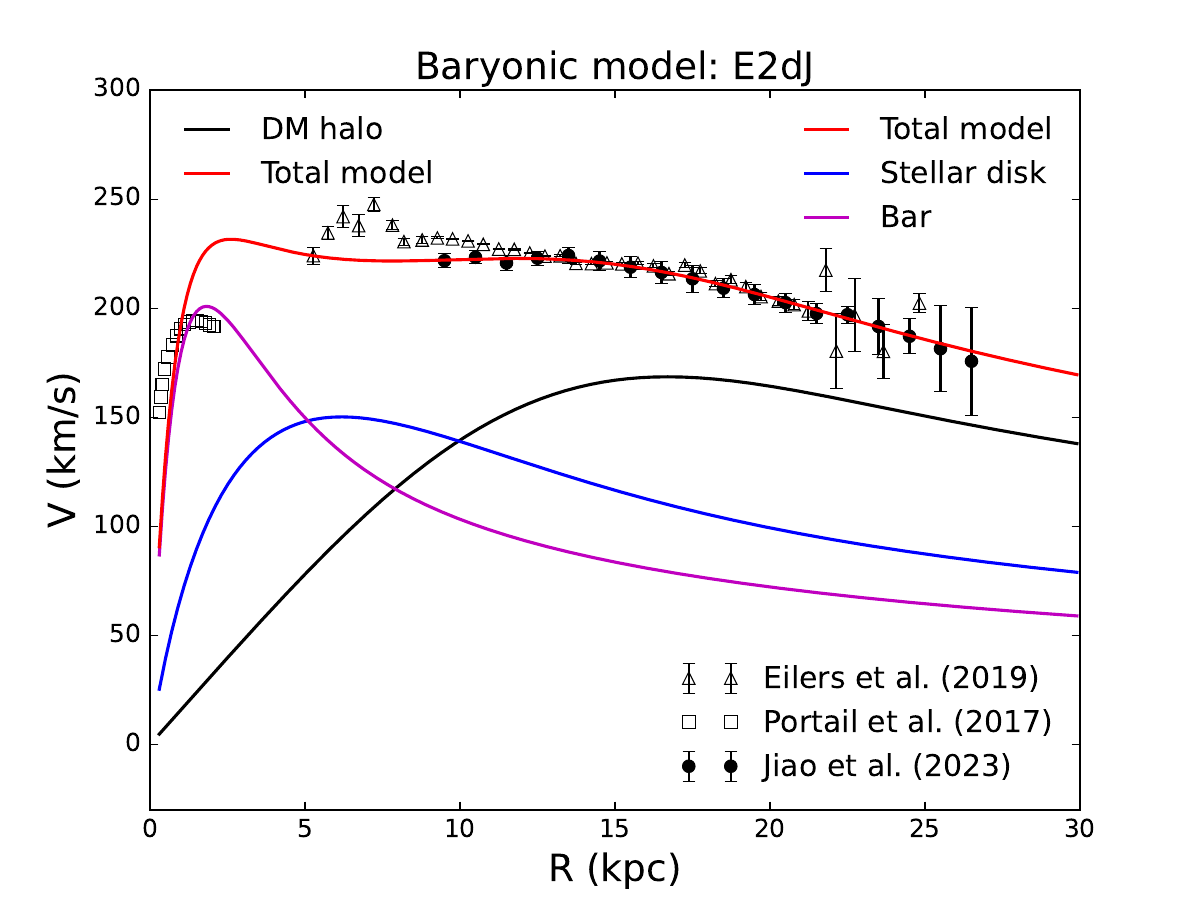}\includegraphics[scale=0.18]{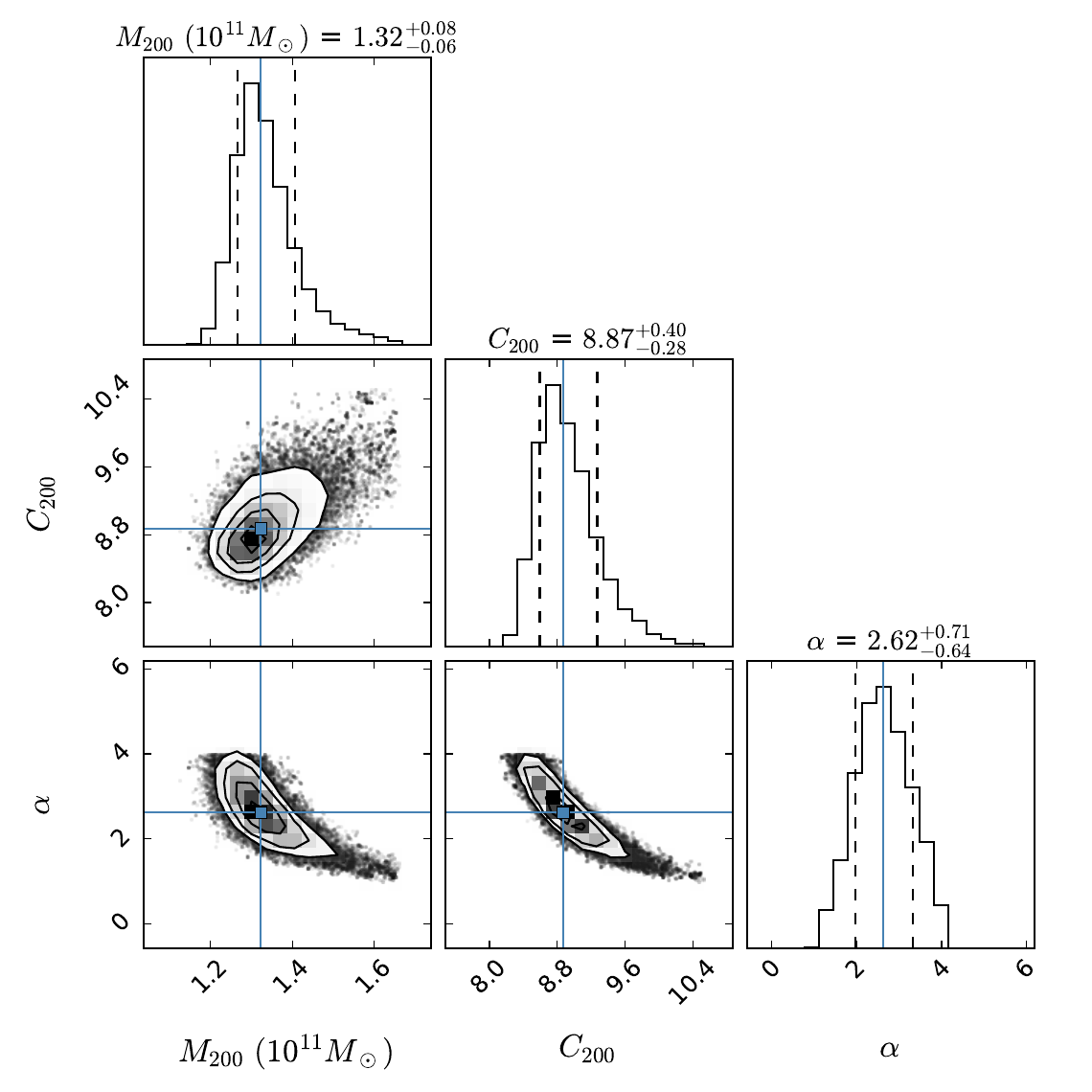}
    \includegraphics[scale=0.25]{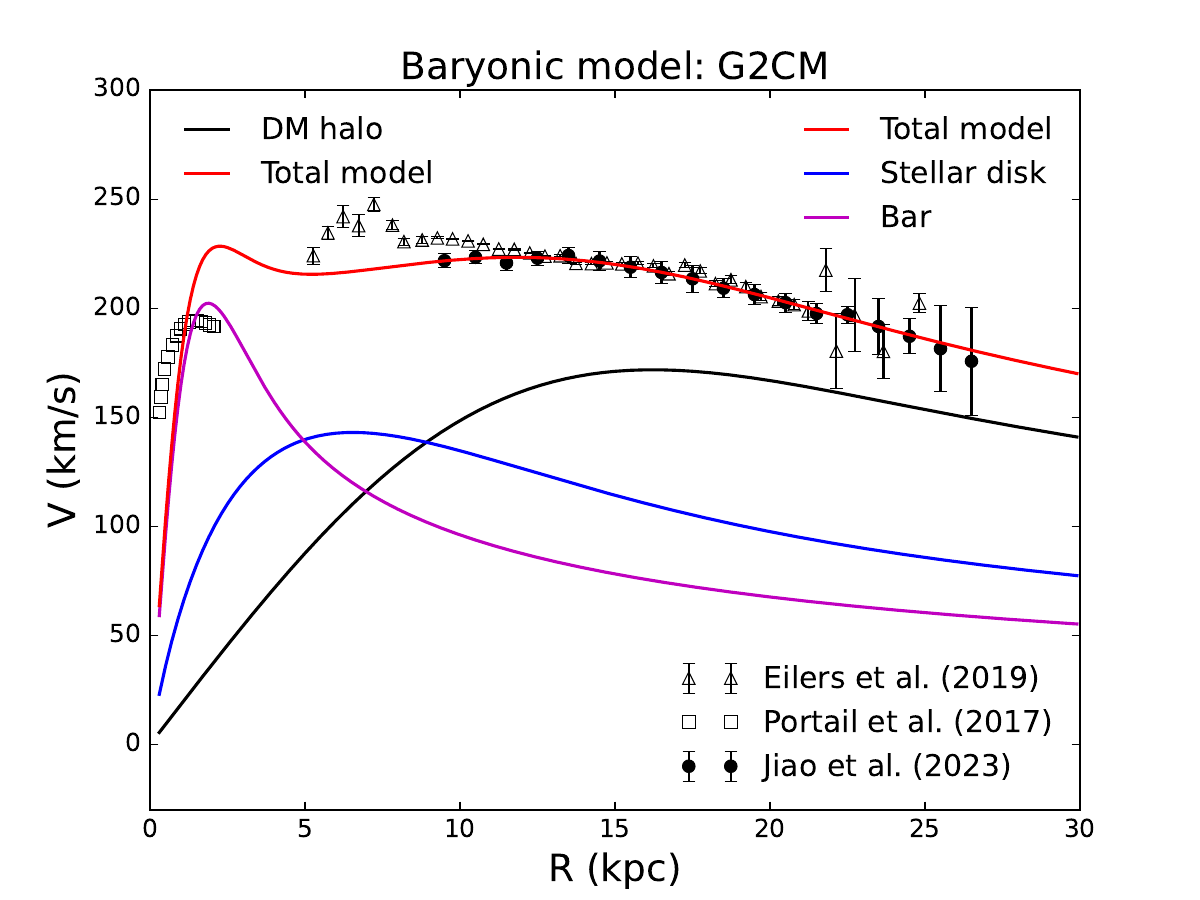}\includegraphics[scale=0.18]{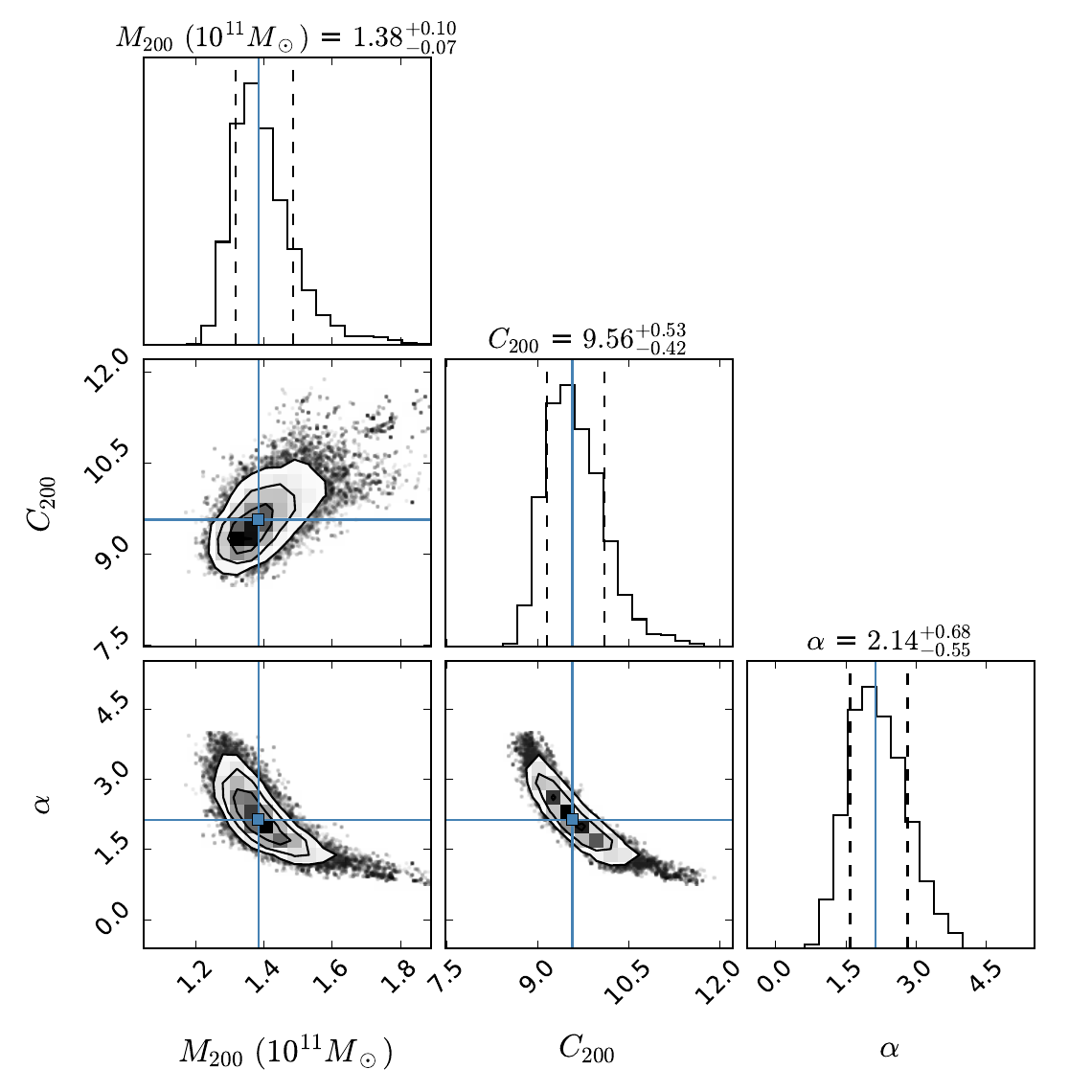}\\
    \includegraphics[scale=0.25]{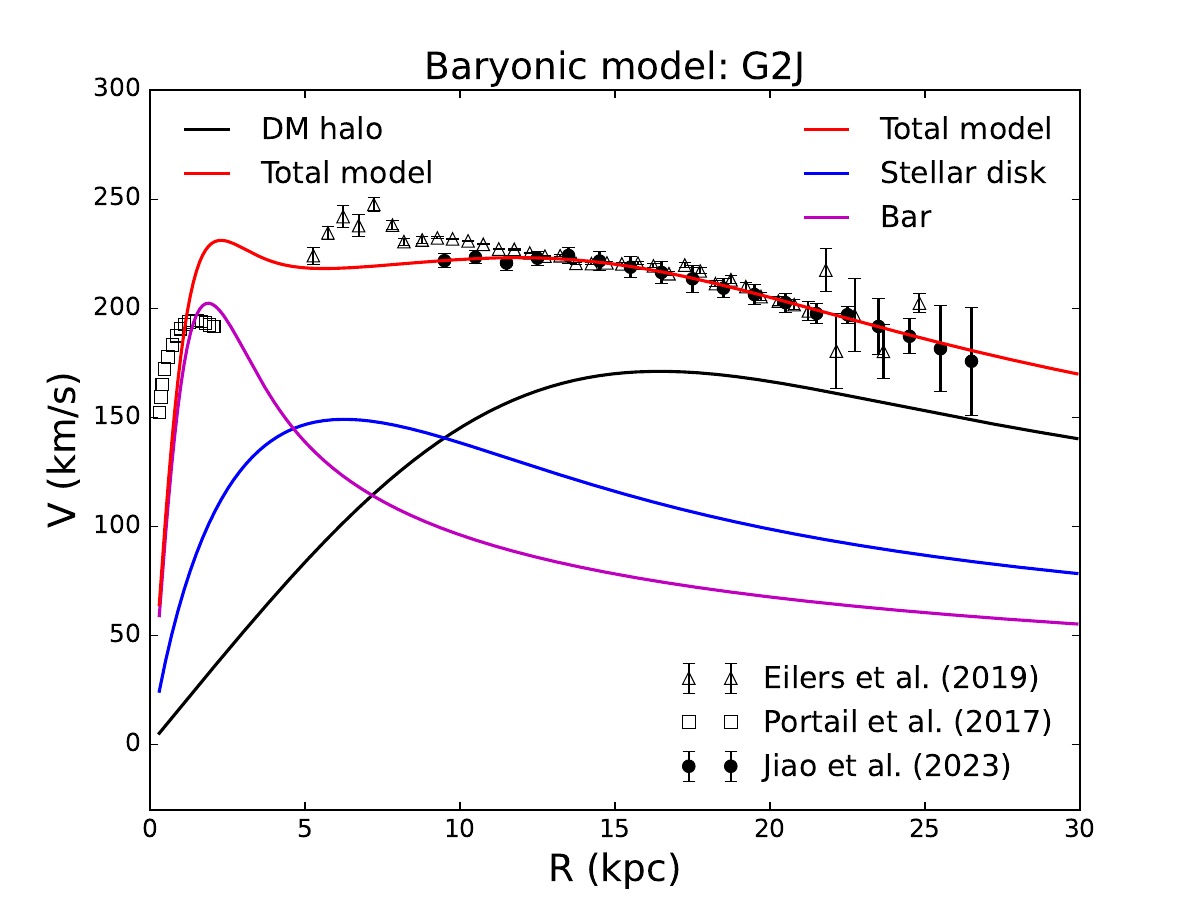}\includegraphics[scale=0.18]{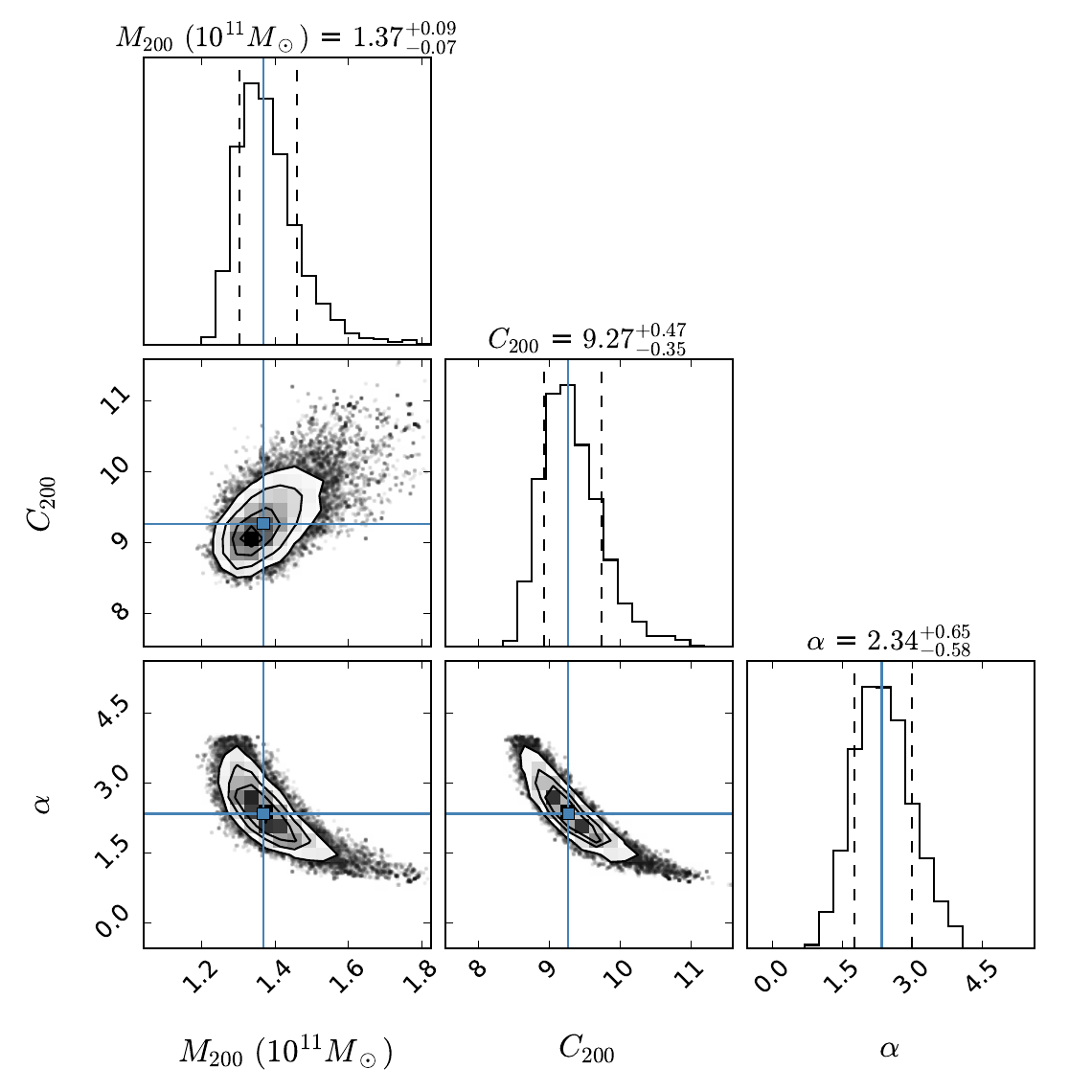}
    \includegraphics[scale=0.25]{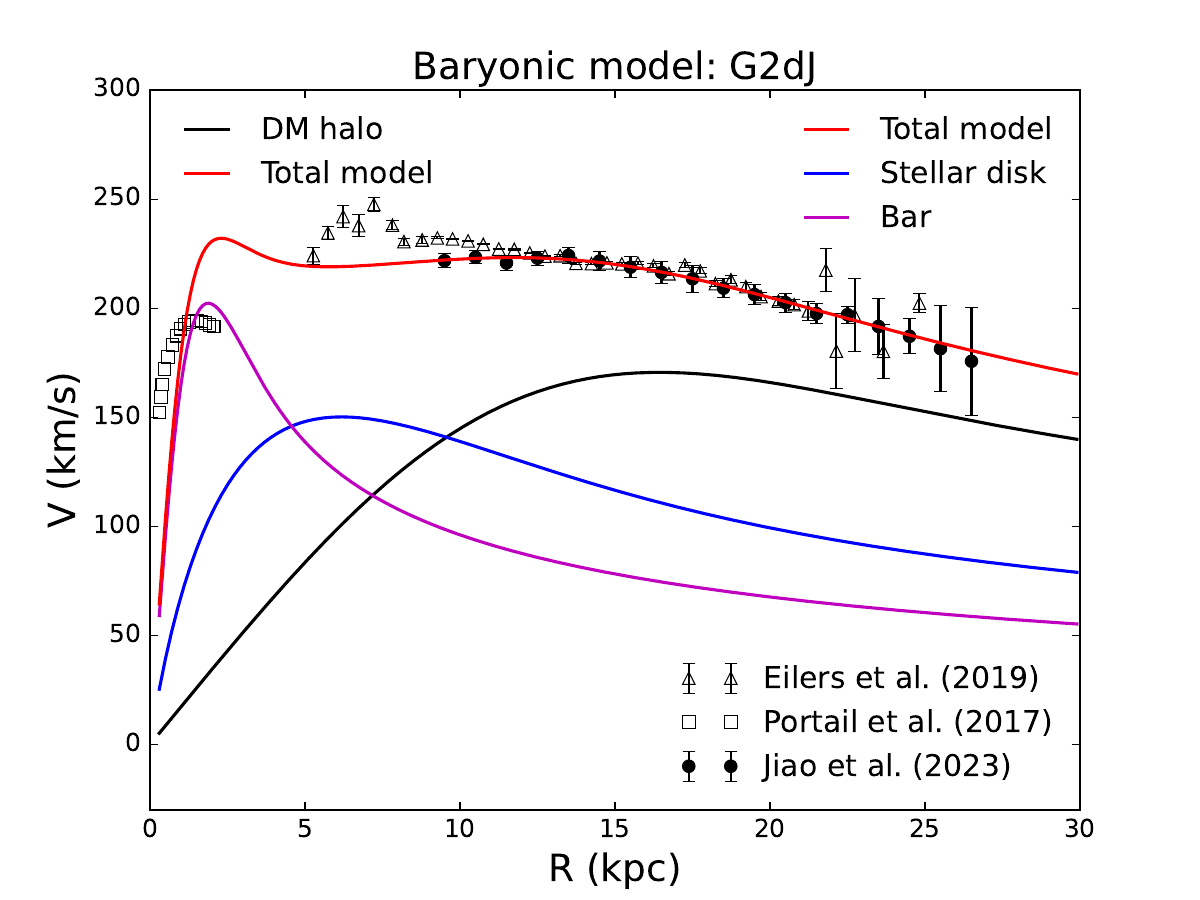}\includegraphics[scale=0.18]{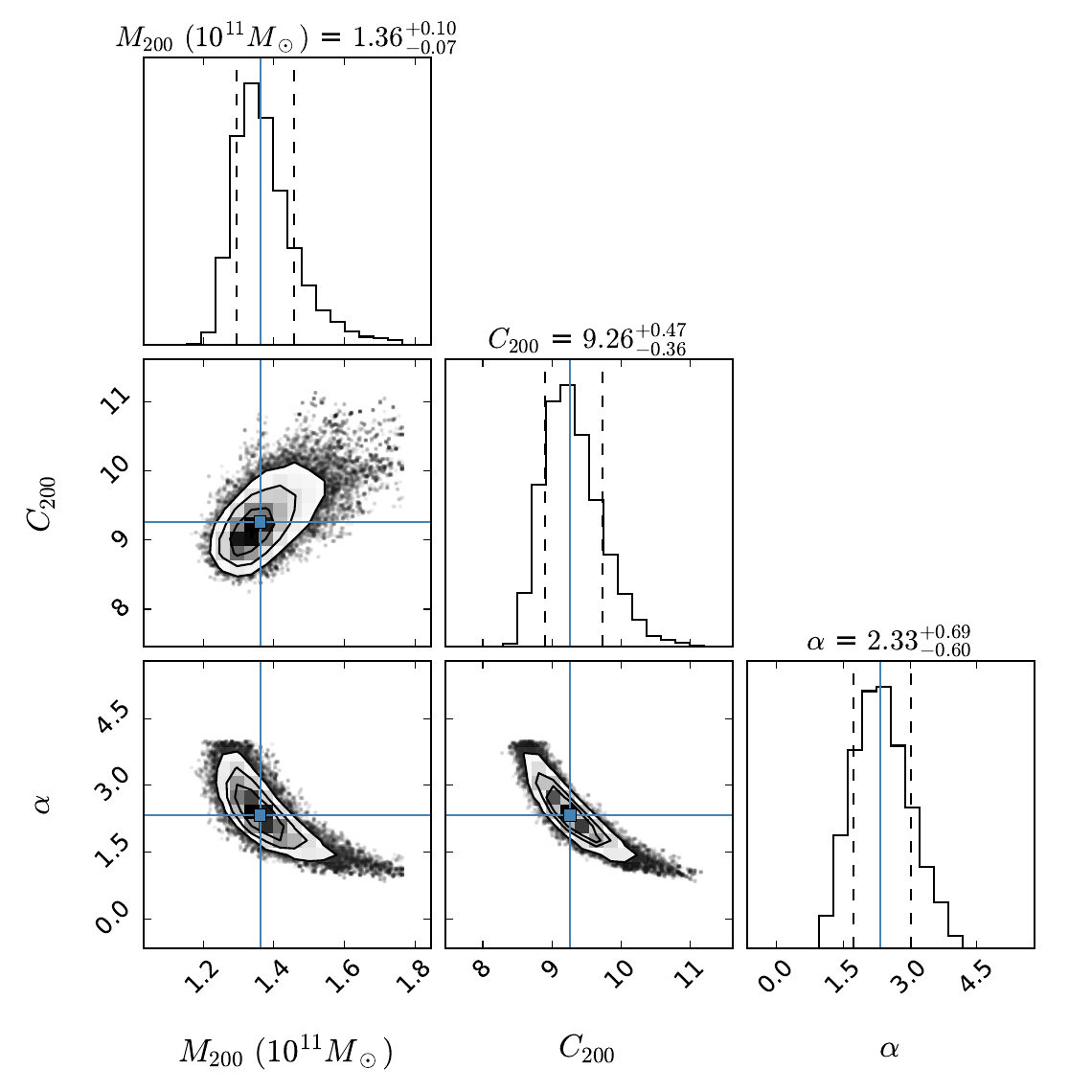}\\
\caption{Fits of Galactic circular velocities using the Einasto model and the posterior distributions of fitting parameters without implementing adiabatic halo contraction using 12 baryonic models.}
\label{fig:fitsNC}
\end{figure*}

\begin{figure*}[b]
\centering
    \includegraphics[scale=0.25]{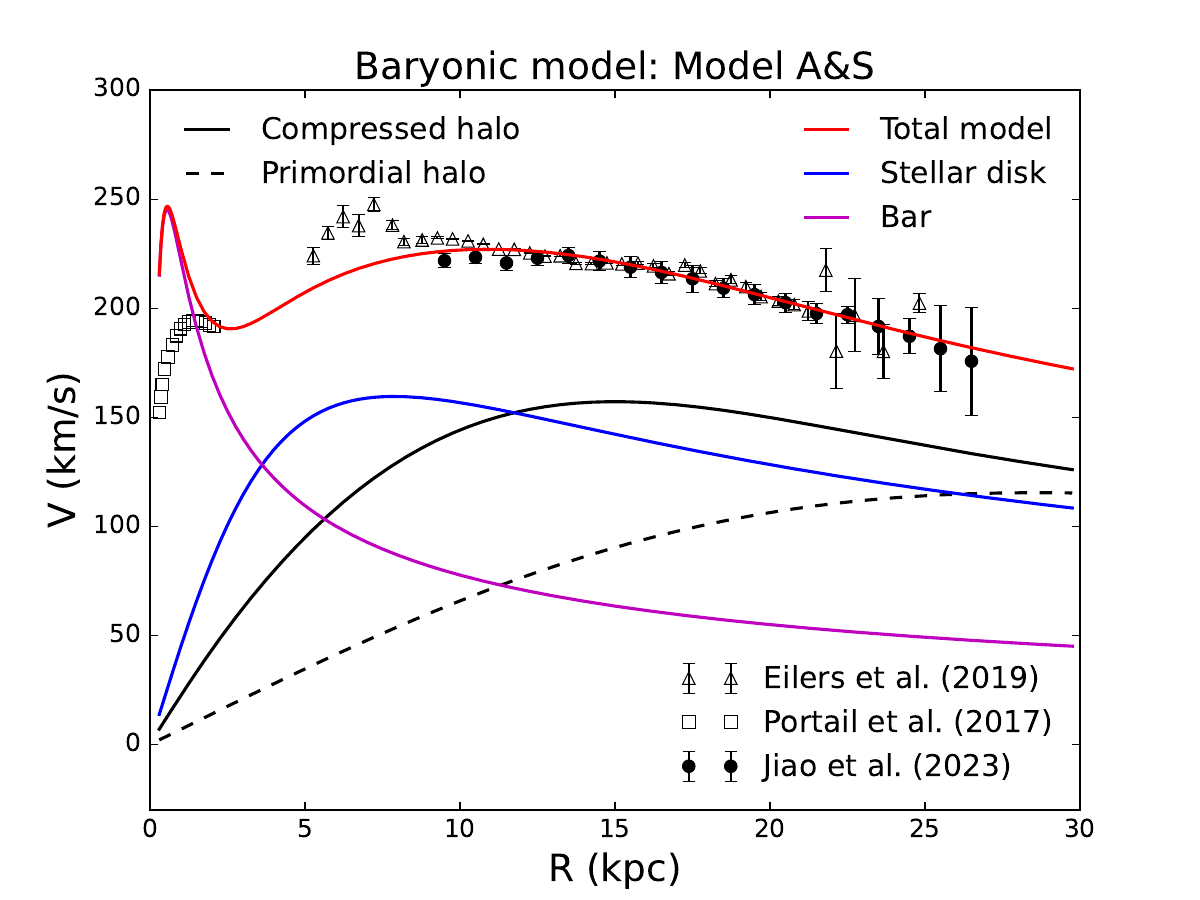}\includegraphics[scale=0.25]{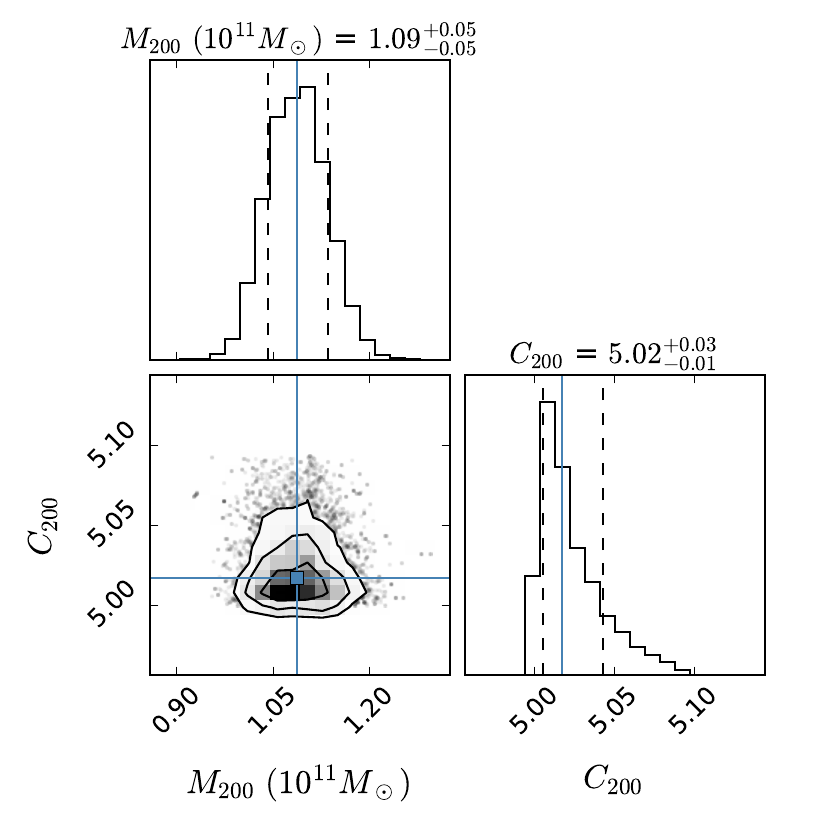}
    \includegraphics[scale=0.25]{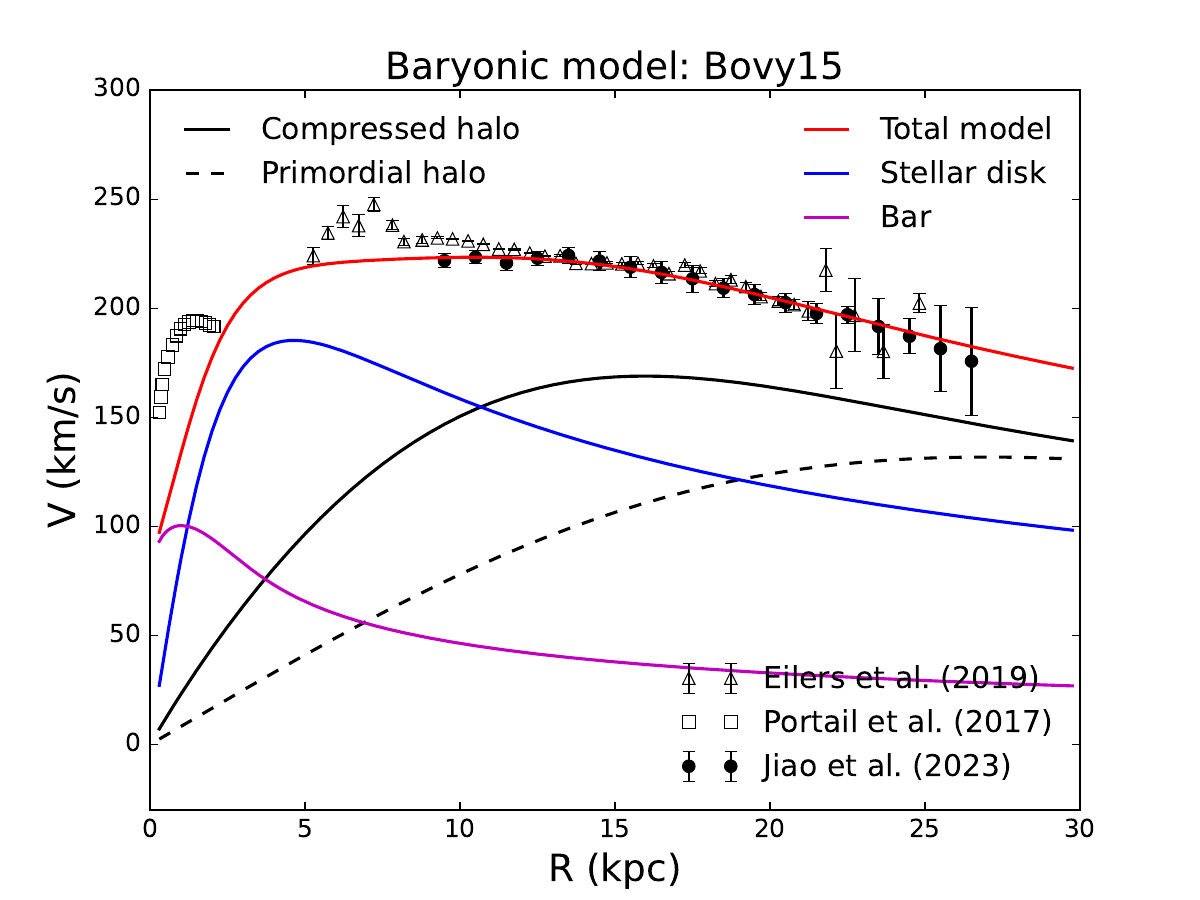}\includegraphics[scale=0.25]{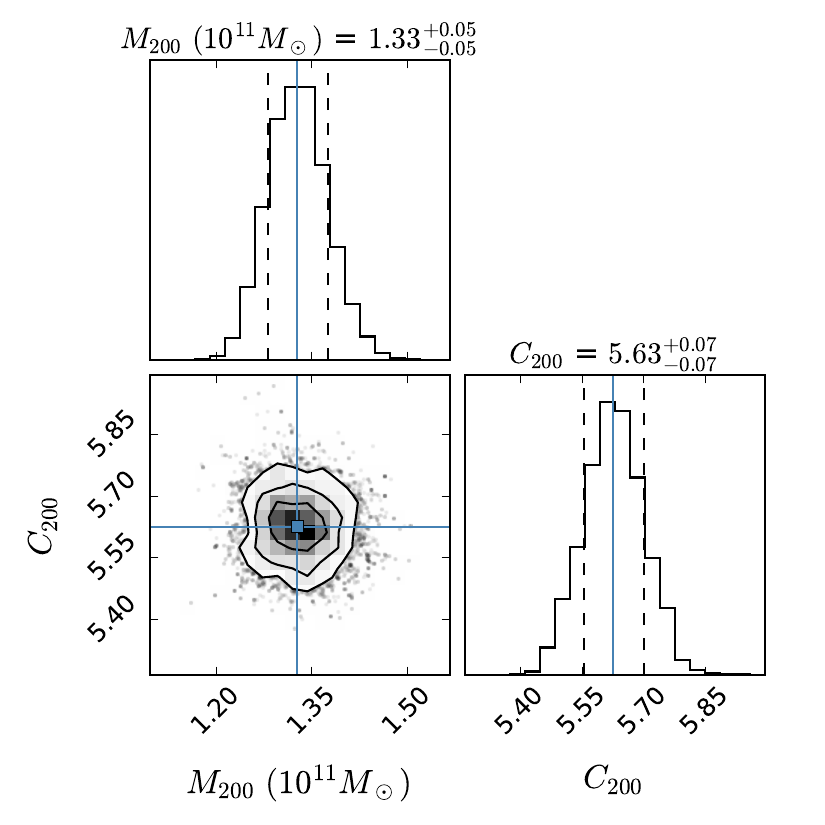}\\
    \includegraphics[scale=0.25]{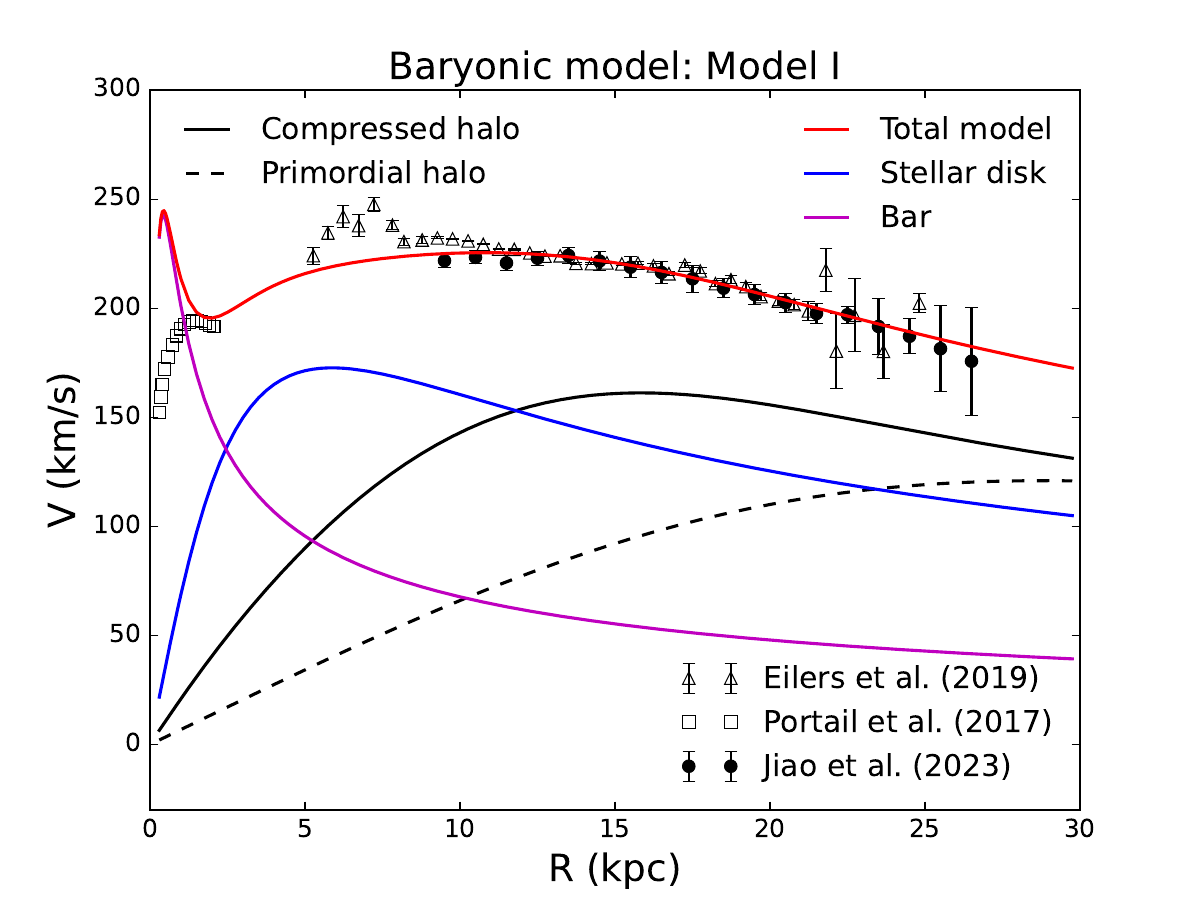}\includegraphics[scale=0.25]{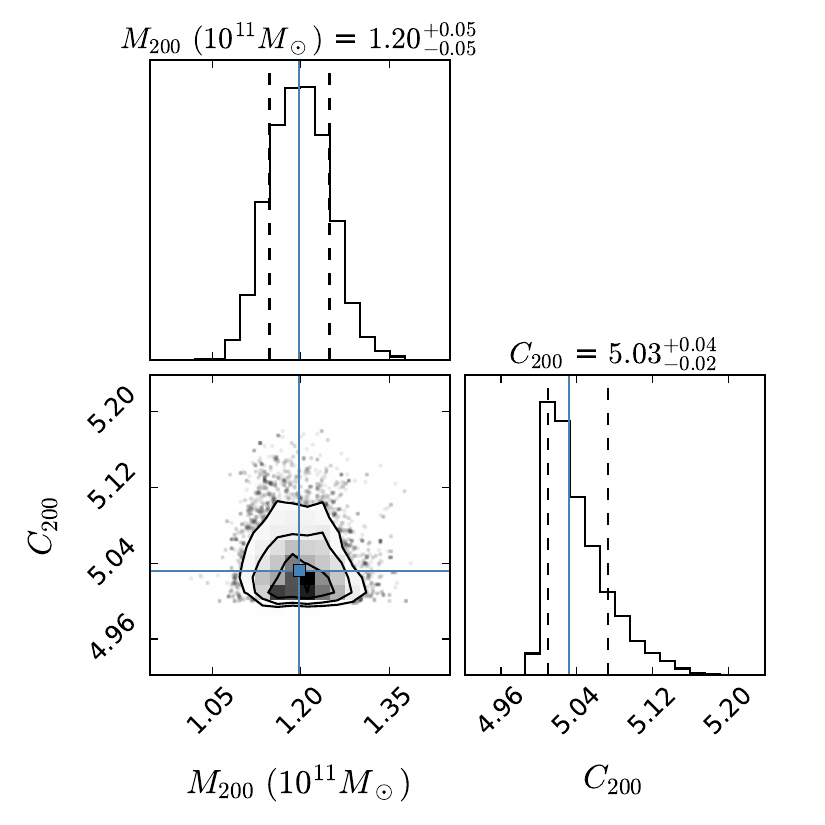}
    \includegraphics[scale=0.25]{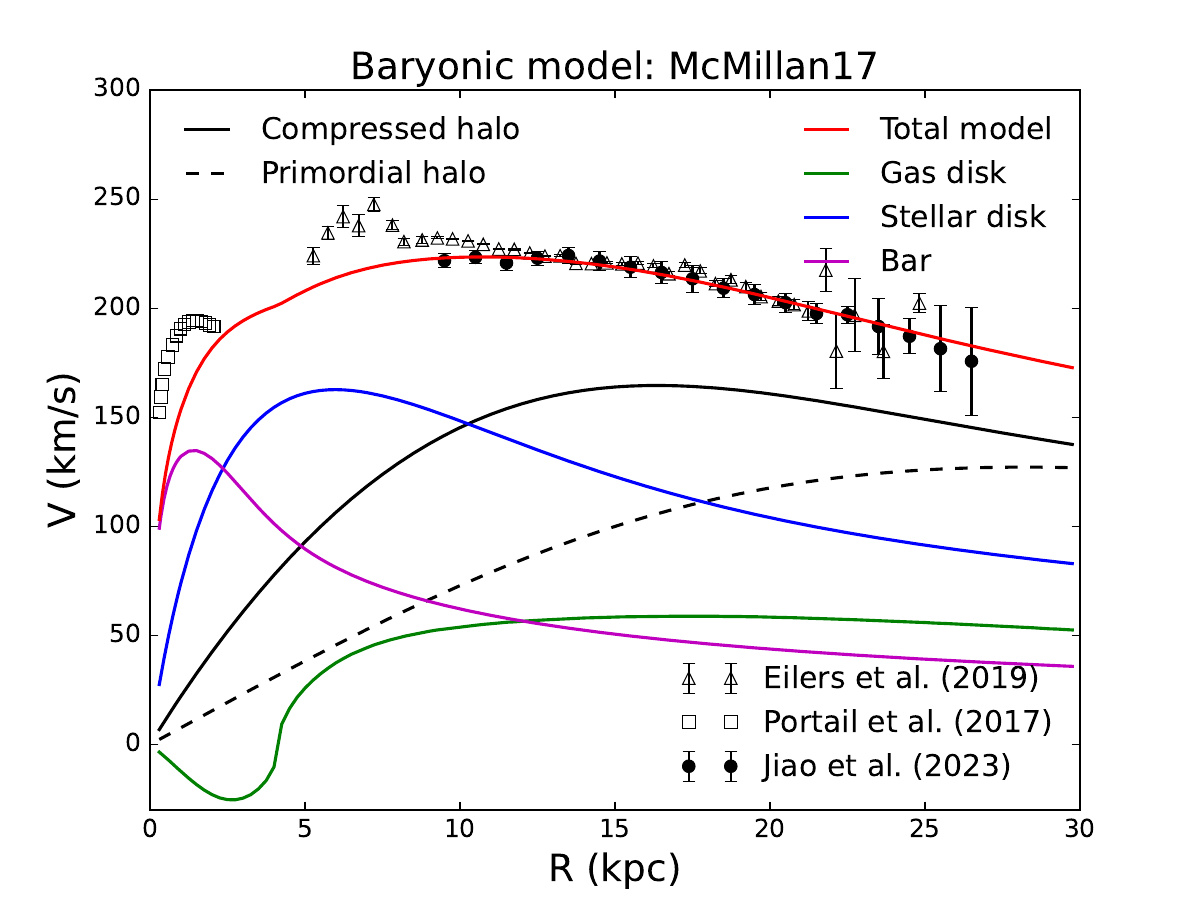}\includegraphics[scale=0.25]{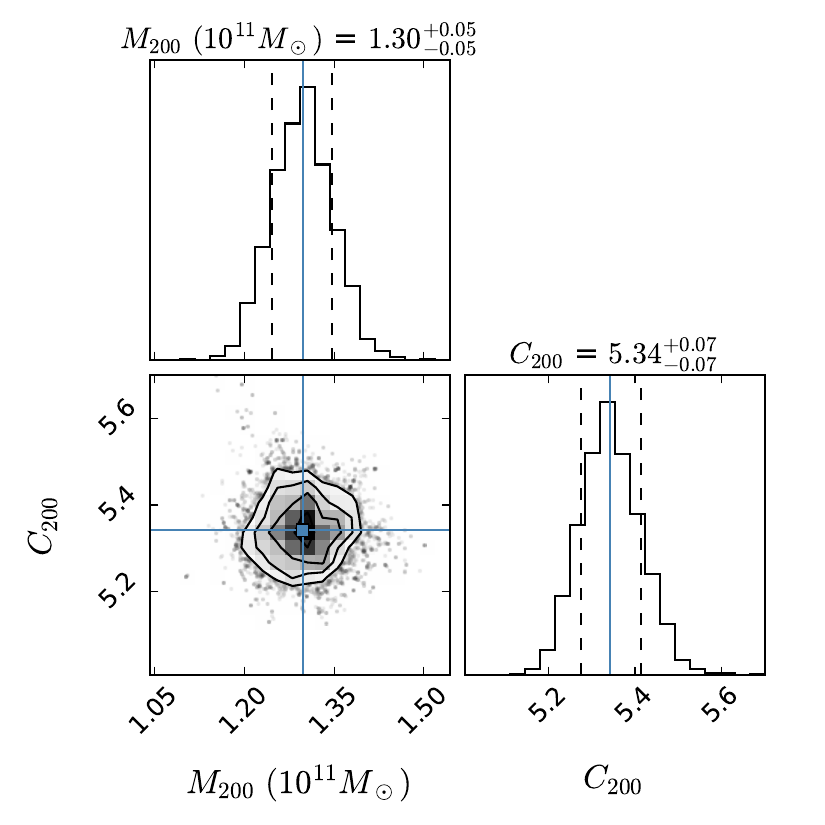}\\
    \includegraphics[scale=0.25]{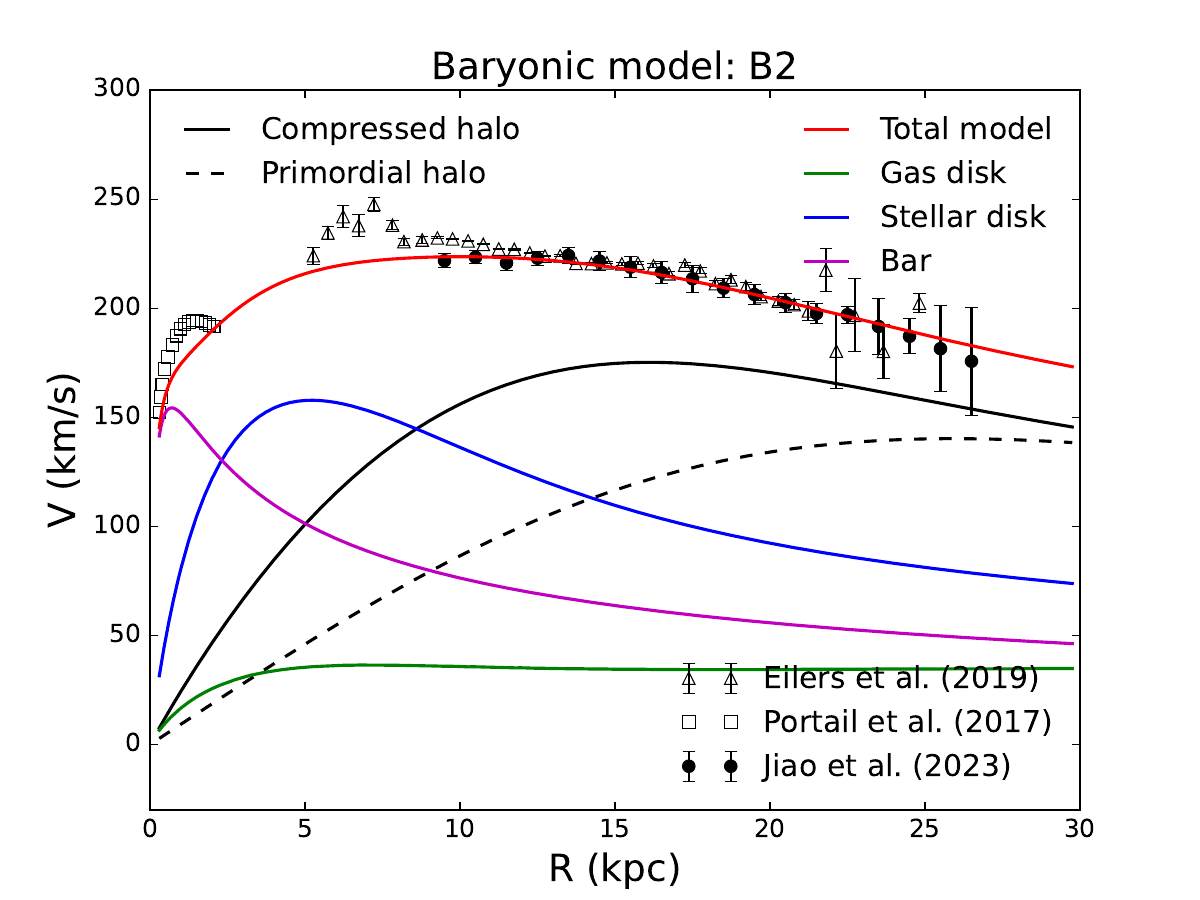}\includegraphics[scale=0.25]{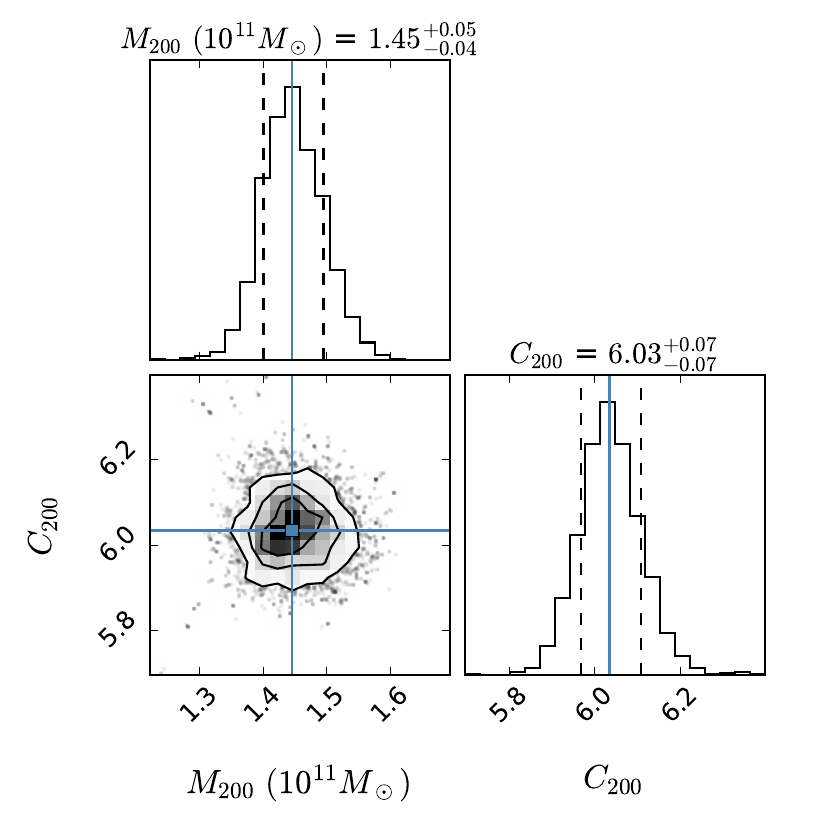}
    \includegraphics[scale=0.25]{Vc_McGaugh19.pdf}\includegraphics[scale=0.25]{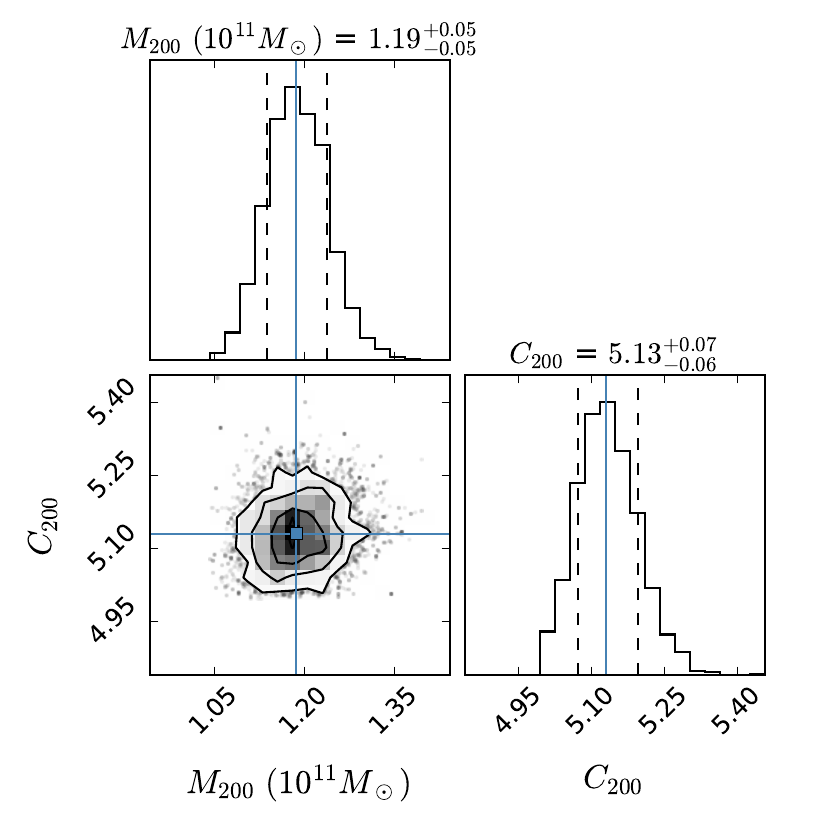}\\
    \includegraphics[scale=0.25]{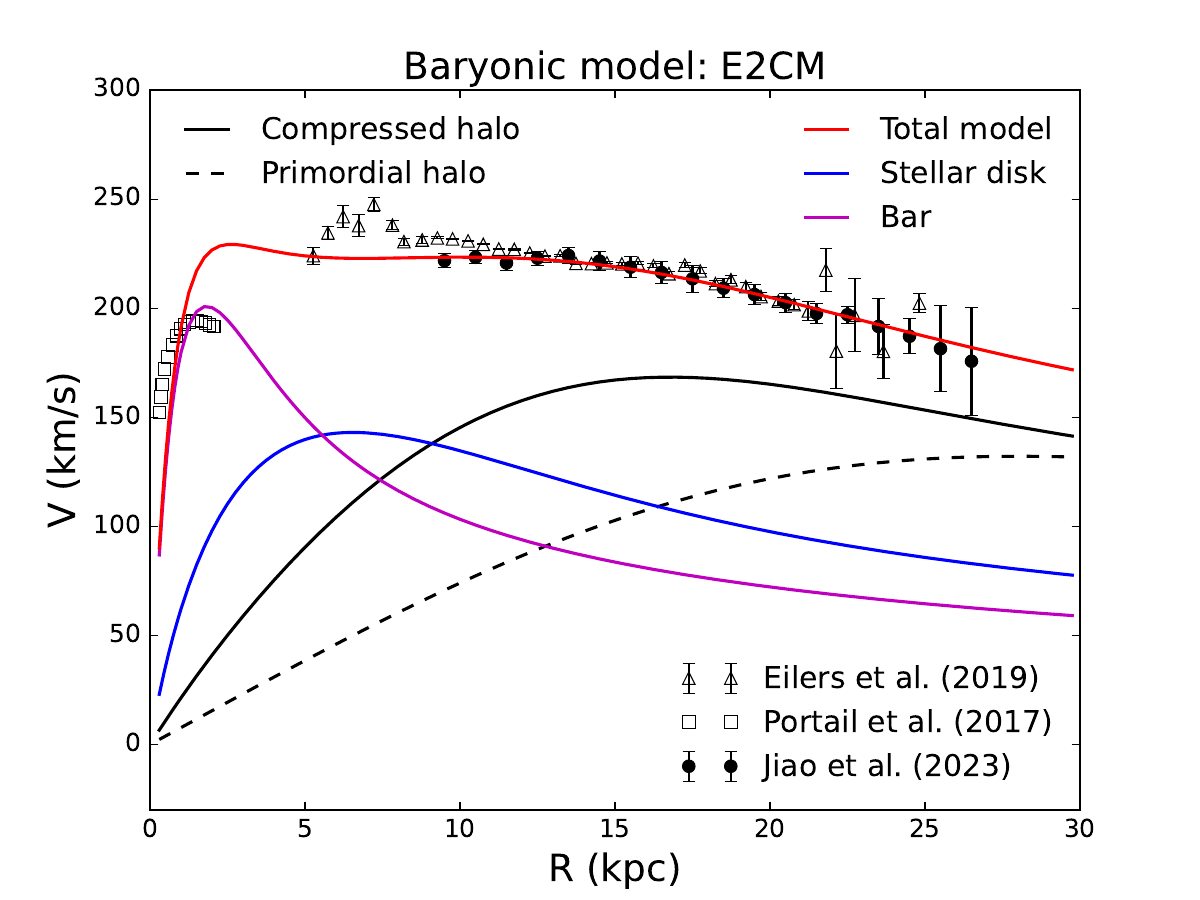}\includegraphics[scale=0.25]{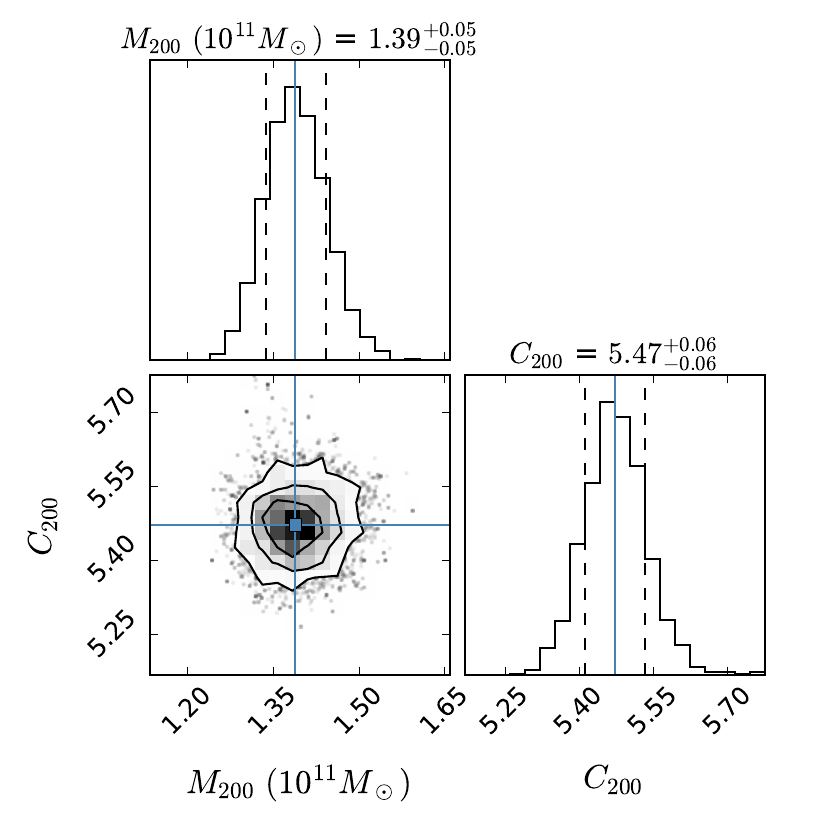}
    \includegraphics[scale=0.25]{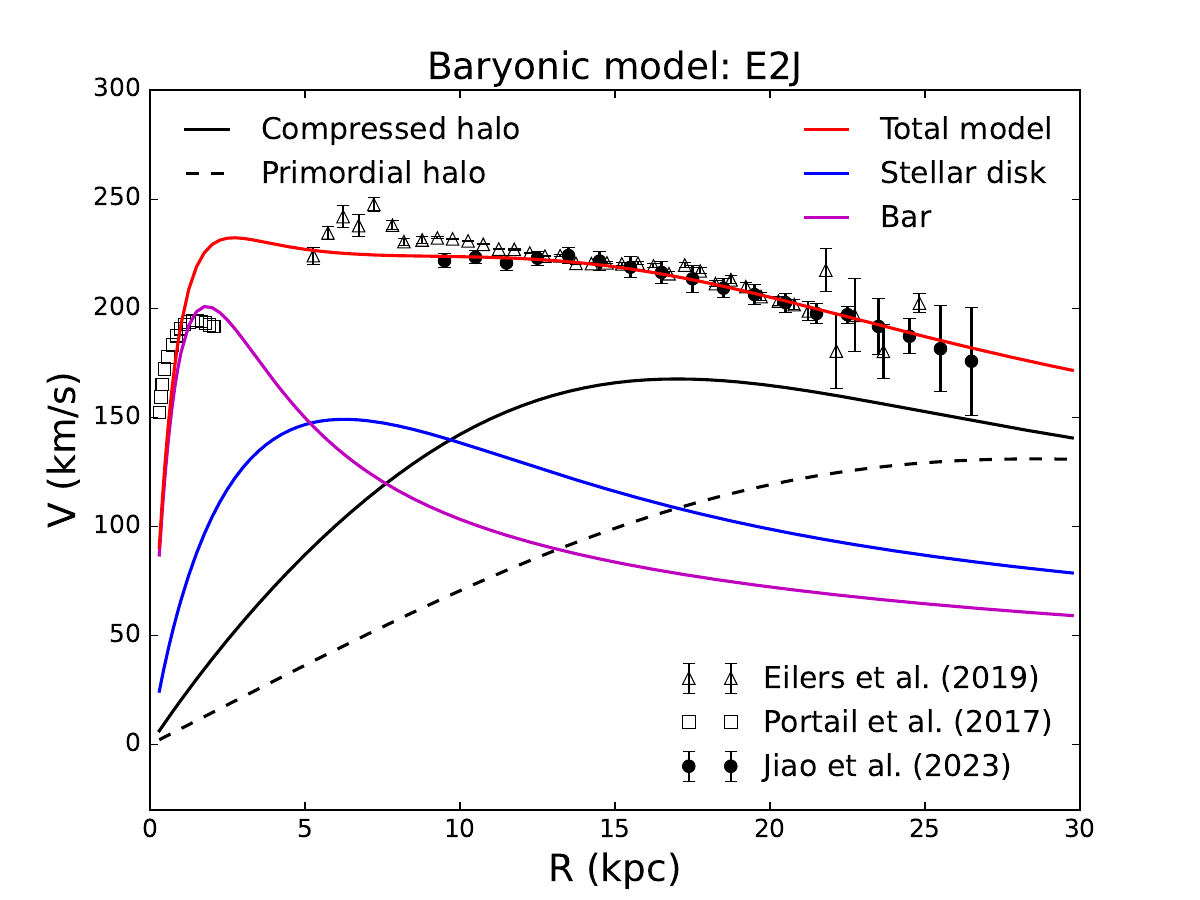}\includegraphics[scale=0.25]{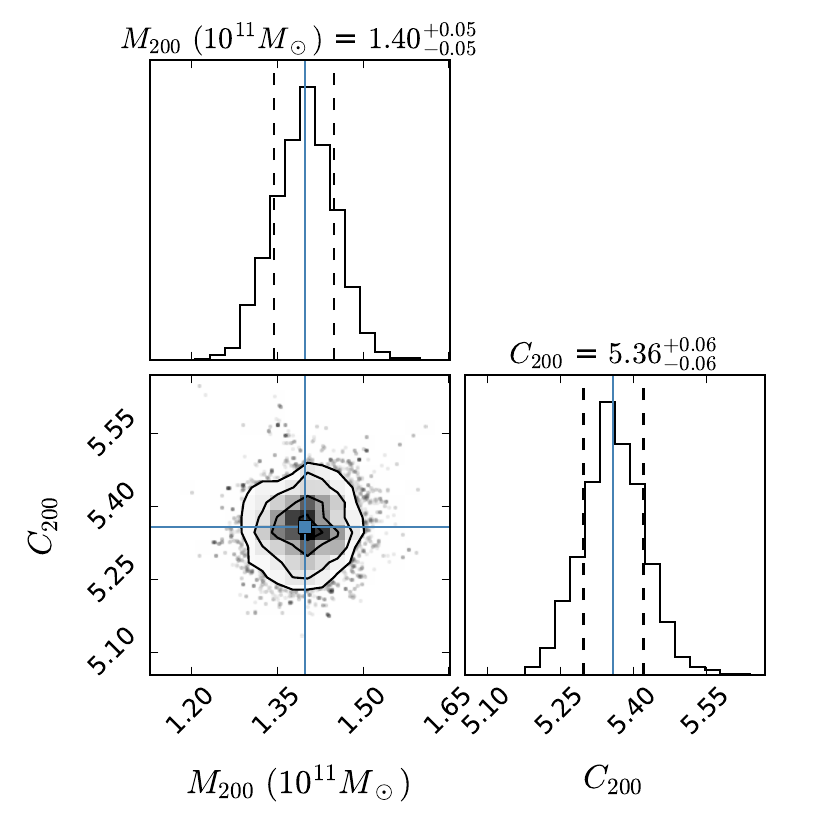}\\
    \includegraphics[scale=0.25]{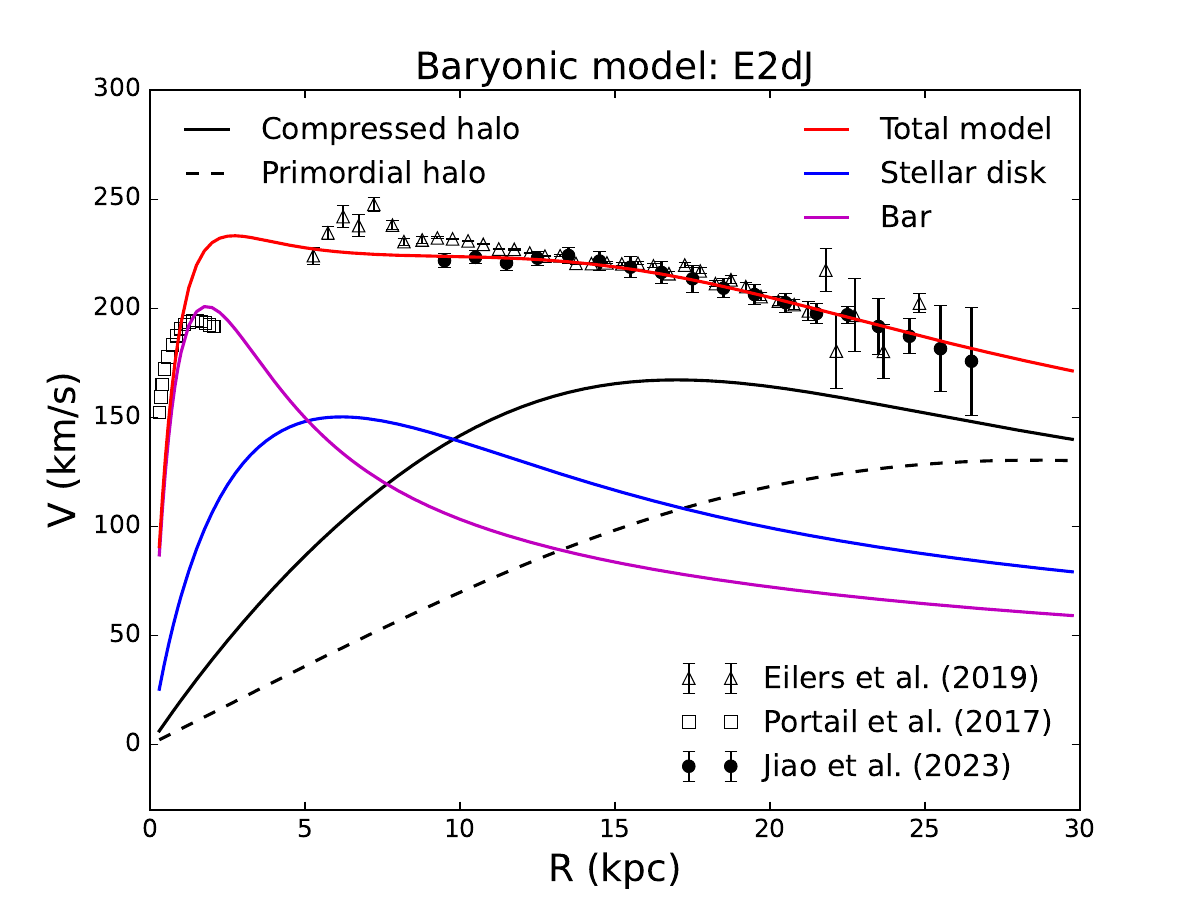}\includegraphics[scale=0.25]{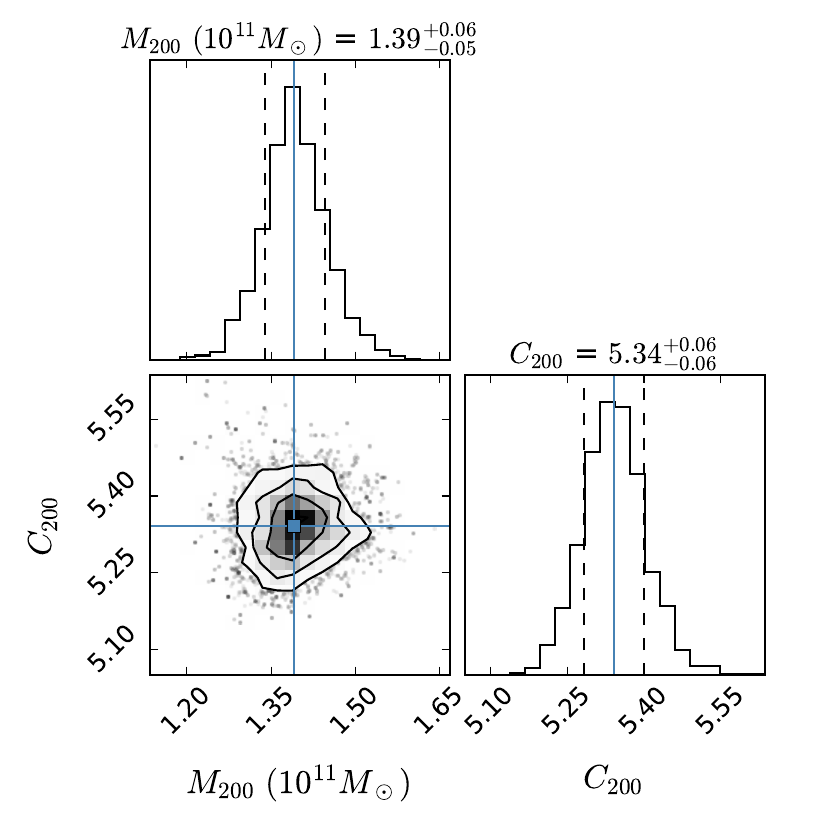}
    \includegraphics[scale=0.25]{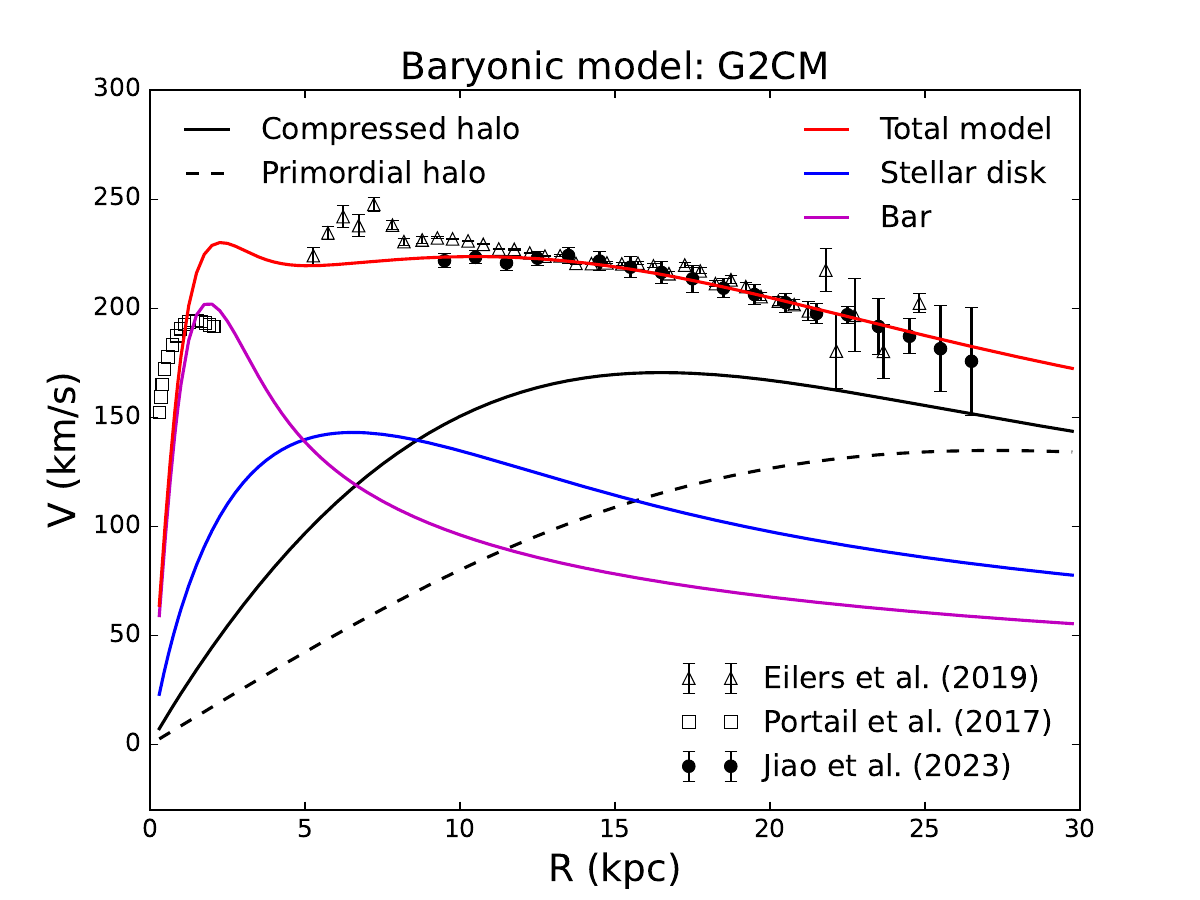}\includegraphics[scale=0.25]{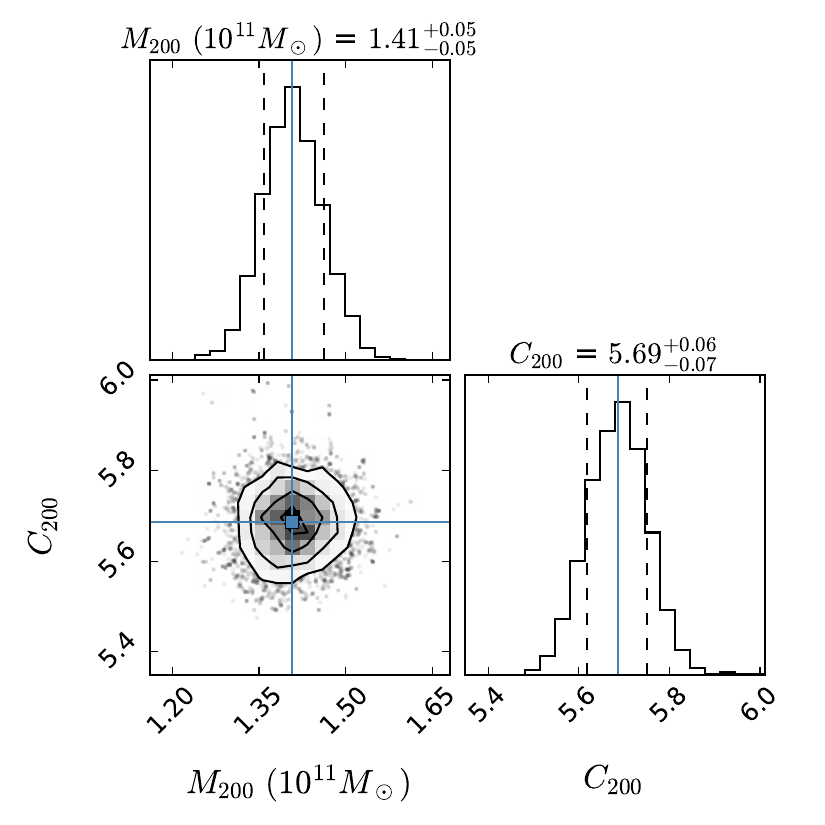}\\
    \includegraphics[scale=0.25]{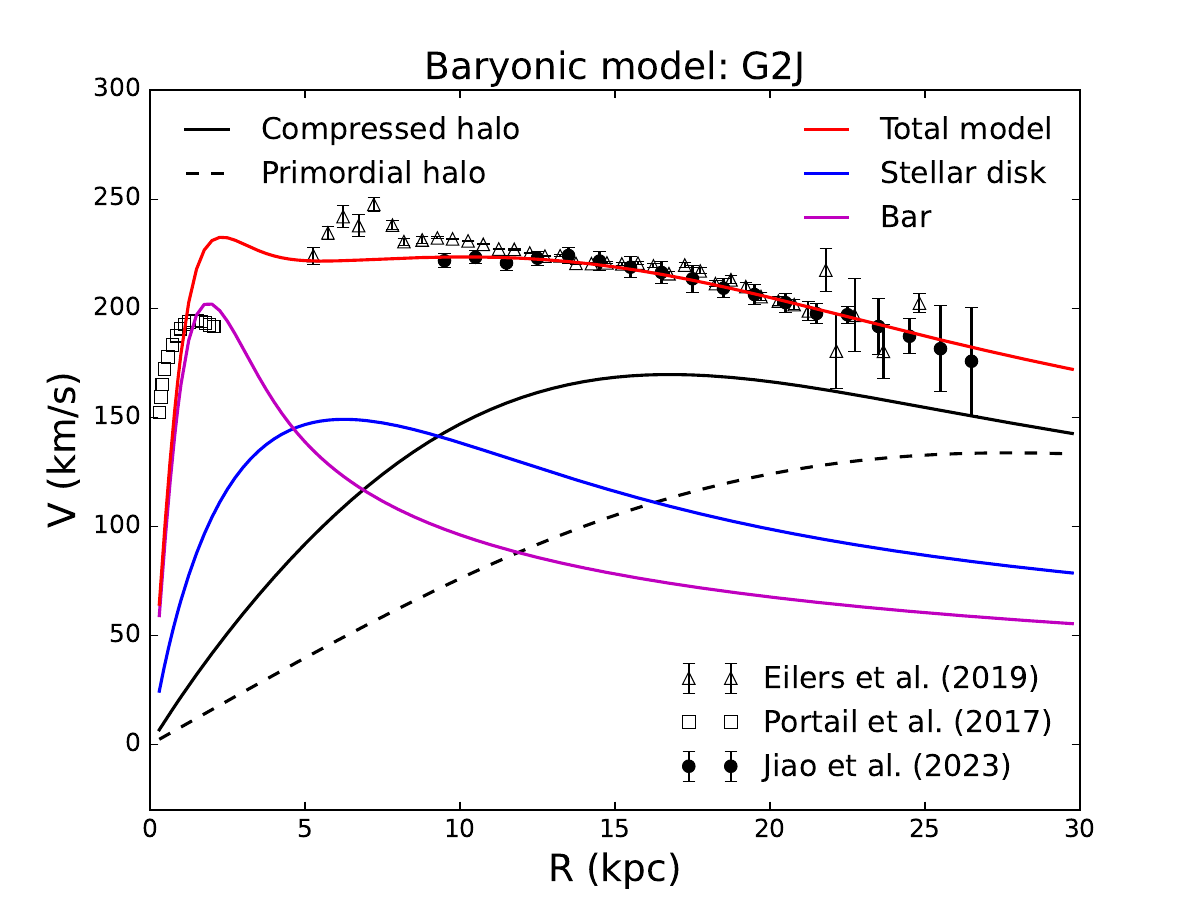}\includegraphics[scale=0.25]{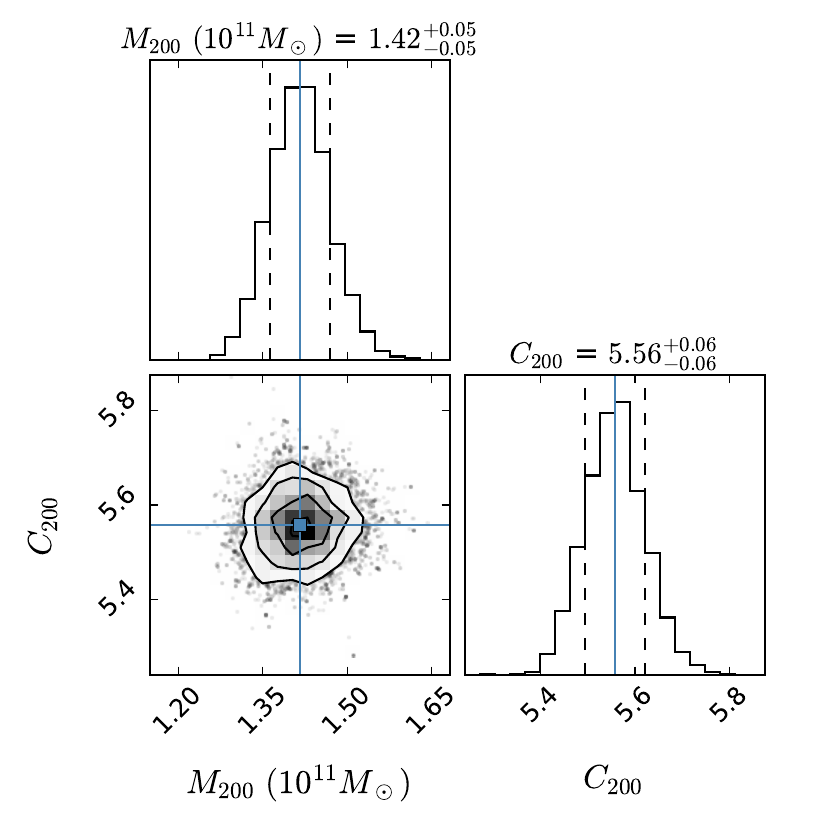}
    \includegraphics[scale=0.25]{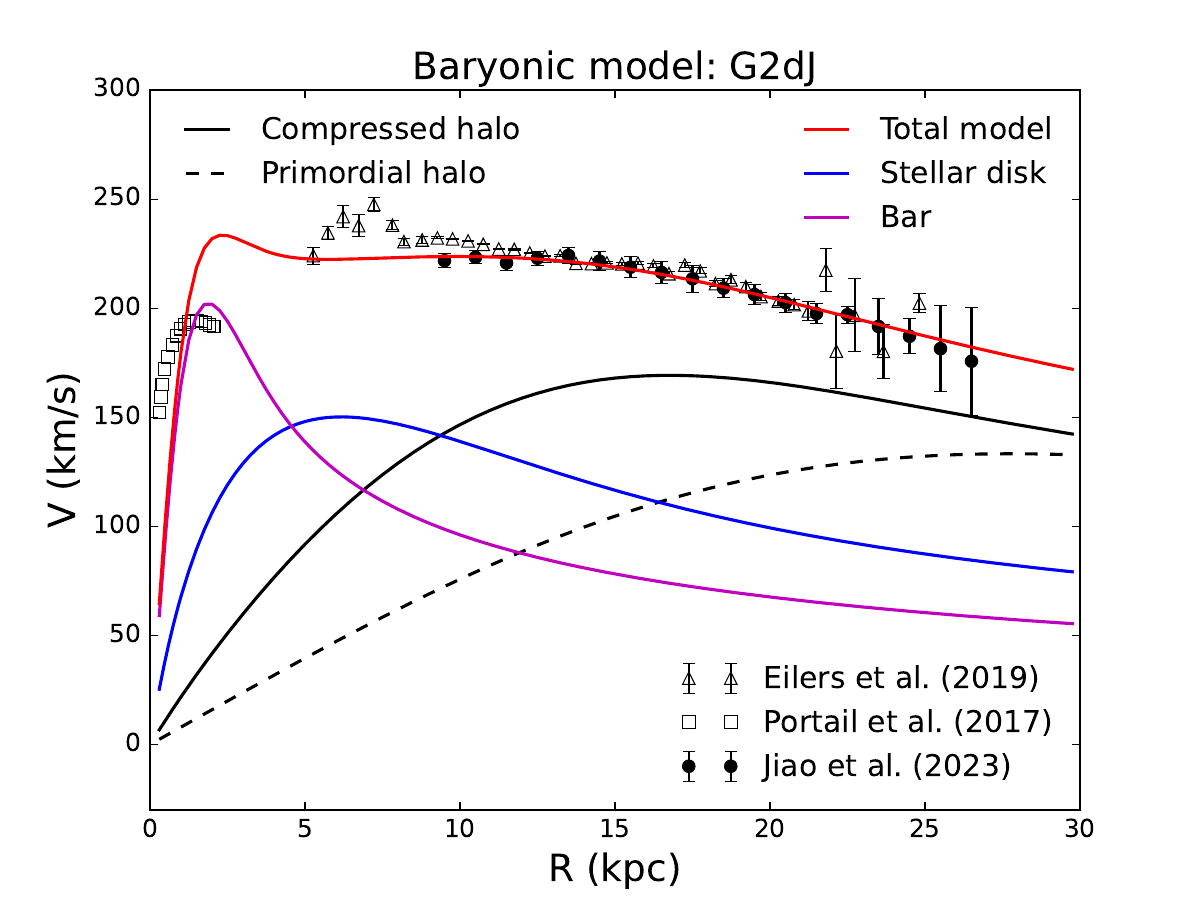}\includegraphics[scale=0.25]{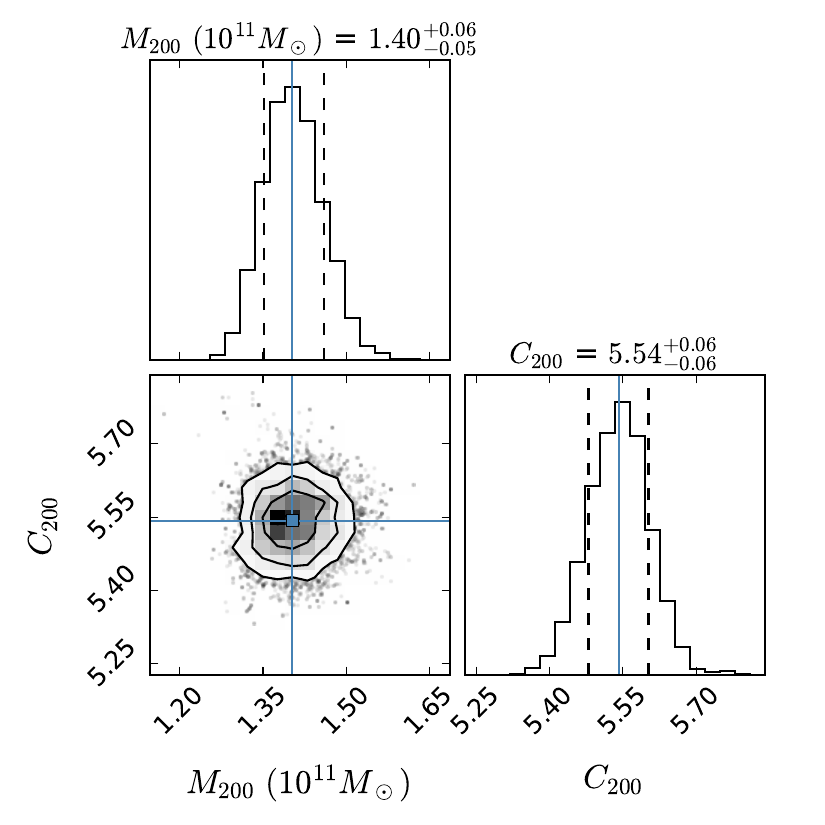}\\
\caption{Fits of Galactic circular velocities using the Einasto model and the posterior distributions of fitting parameters implementing adiabatic halo contraction using 12 baryonic models.}
\label{fig:fits}
\end{figure*}

We use \textsc{emcee} \citep{Foreman-Mackey2013} to map the posterior distributions of the fitting parameters and impose flat priors within the following hard boundaries: $0.01<\alpha<4.0$, $5.0<C_{200}<30.0$ and $50<V_{200}<200$ km/s. We checked and verified these ranges are sufficiently wide. For comparison, we fit the circular velocities without implementing adiabatic contraction of dark matter halos. The fit results and posterior distributions of the free parameters are shown in Fig.~\ref{fig:fitsNC}. When implementing baryonic compression, the computational time increased dramatically. For the Einasto model, one has to generate different input tables for different values of the shape parameter $\alpha$. This would require significantly more computational time. For simplicity, we fix the value of $\alpha$ for each baryonic model using the fitting results without implementing halo contraction. This simplification is supported by the fact that the declining part of the circular velocity profile from \textit{Gaia} strongly constrains the outer density slope of the current halo, which is insensitive to baryonic compression. As such, including halo contraction or not does not significantly influence the value of $\alpha$. The decent fits in Fig.~\ref{fig:fits} further justify the simplification.

\begin{figure*}[b]
\centering
\includegraphics[scale=0.35]{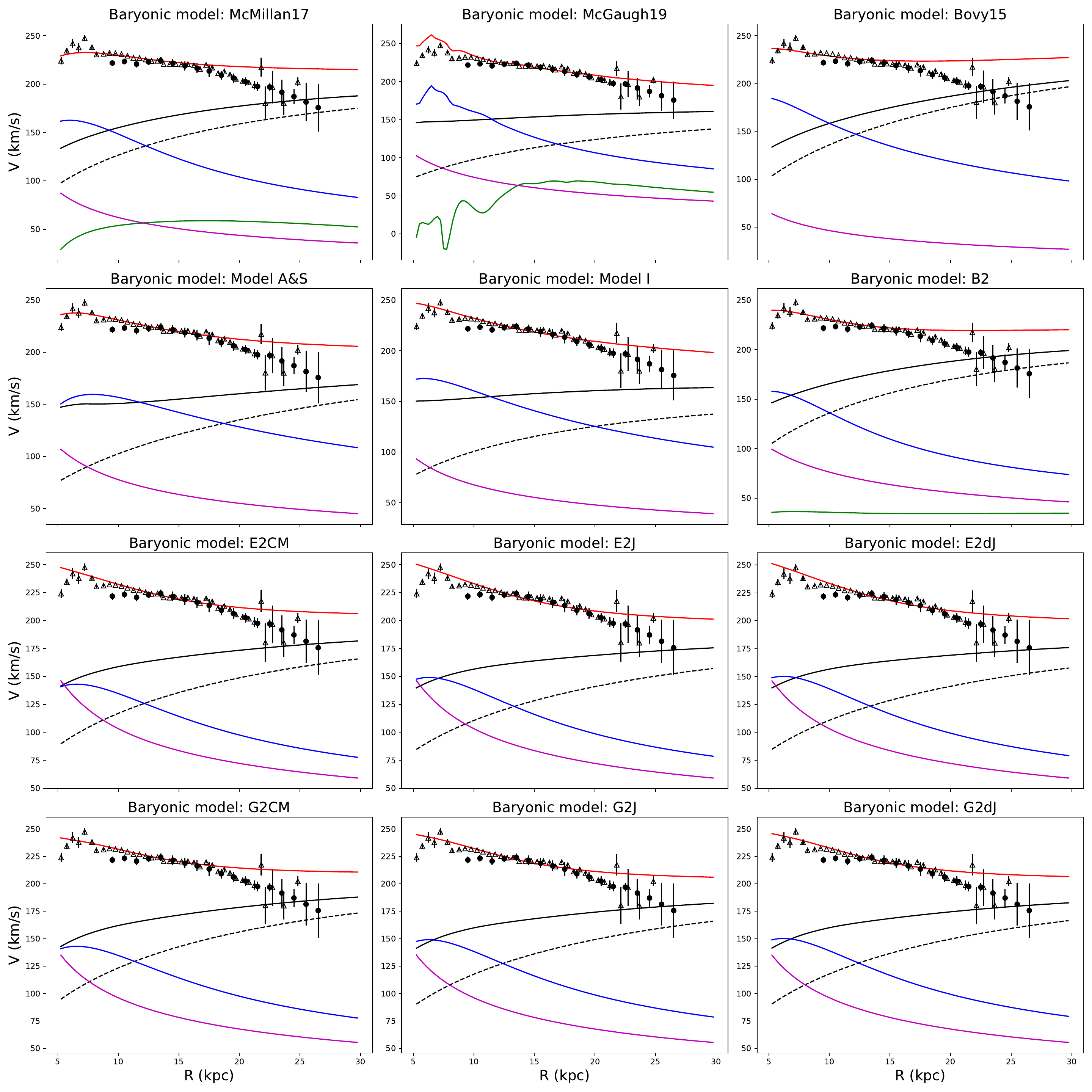}
\caption{Fits of Galactic circular velocities using the NFW model implementing adiabatic halo contraction using 12 baryonic models. Data points with errors are the rotation velocities from \cite{Jiao2023}, while open triangles show the data from \cite{Eilers2019}, which are not fitted. Blue, purple, green and black solid lines correspond to the contributions by the stellar disk, central bar, gas (and dust if any), and compressed dark matter halo, respectively. The total contributions are shown using red solid lines. Black dashed lines are the inferred primordial halos.}
\label{fig:NFWfits}
\end{figure*}
    
\end{appendix}
\end{document}